\documentclass[12pt]{article}

\usepackage{graphicx}
\usepackage{amsfonts}
\usepackage{amssymb}
\usepackage{amsbsy}

\input epsf.sty

\newcommand{\BEQ}{\begin{equation}}     % Gleichungen Anfang ..
\newcommand{\BEA}{\begin{eqnarray}}
\newcommand{\EEQ}{\end{equation}}       % .. und Ende
\newcommand{\EEA}{\end{eqnarray}}
\newcommand{\sig}{\sigma}               % sigma
\newcommand{\ups}{\upsilon}             % upsilon
\newcommand{\D}{{\rm d}}                % gerades d fuer Ableitungen
\newcommand{\II}{{\rm i}}               % gerades i fuer komplexe Einheit
\renewcommand{\Re}{{\rm Re\ }}          % Realteil
\newcommand{\wit}[1]{\widetilde{#1}}    % weite Schlange
\newcommand{\ket}[1]{\left|#1\right\rangle}      % Ket-Zustand
\newcommand{\bra}[1]{\left\langle #1\right|}     % Bra-Zustand
\newcommand{\braket}[2]{\left\langle\left. #1\right| #2 \right\rangle} 
\newcommand{\vm}[1]{\check{#1}}                  % Hut verkehrt fuer Math.
\newcommand{\rar}{\rightarrow}          % Pfeil nach rechts

\renewcommand{\vec}[1]{\boldsymbol{#1}} % Vektoren fettgedruckt

\newcommand{\zeile}[1]{\vskip #1 \baselineskip} % N Zeilen ueberschlagen
\newcommand{\vekz}[2]
     {\mbox{${\begin{array}{c} #1  \\ #2 \end{array}}$}}
\newcommand{\matz}[4]
     {\mbox{${\begin{array}{cc} #1 & #2  \\ #3 & #4 \end{array}}$}}
\begin{document}

%%%%%%%%%%%%%%%%%%%%%%%%%%%%%%%%%%%%%%%%%%%%%%%%%%%%%%%%%%%%%%%%%%%%%%%%%%%%%%%%
%       Here begins the titlepage, you may edit this as you wish               %
%%%%%%%%%%%%%%%%%%%%%%%%%%%%%%%%%%%%%%%%%%%%%%%%%%%%%%%%%%%%%%%%%%%%%%%%%%%%%%%%
\begin{titlepage}

%%{\hfill \tt \today; pr\'eliminaire !}

\vskip 1.5 cm
\begin{center}
{\Large \bf Reaction-diffusion processes and their connection with 
integrable quantum spin chains}\footnote{to be published in: A. Kundu
(\'ed) {\it Classical and Quantum Nonlinear Integrable Systems: 
Theory and Applications}, Institute of Physics Publishing (Bristol 2003) \\
ISBN 07503 09598}
 
\end{center}

\vskip 2.0 cm
   
\centerline{ {\bf Malte Henkel} }
\vskip 0.5 cm
\centerline {Laboratoire de Physique des 
Mat\'eriaux (CNRS UMR 7556) et}
\centerline{
Laboratoire Europ\'een de Recherche Universitaire Sarre-Lorraine,}
\centerline{ 
Universit\'e Henri Poincar\'e Nancy I, B.P. 239,} 
\centerline{ 
F -- 54506 Vand{\oe}uvre l\`es Nancy Cedex, France}

\begin{abstract}
\noindent 
This is a pedagogical account on reaction-diffusion systems and their 
relationship with integrable quantum spin chains.

Reaction-diffusion systems are paradigmatic examples of 
non-equi\-li\-brium systems. Their long-time behaviour is 
strongly influenced through fluctuation effects
in low dimensions which renders the habitual mean-field kinetic equations
inapplicable. Starting from the master equation rewritten as a Schr\"odinger
equation with imaginary time, the associated quantum Hamiltonian of certain
one-dimensional reaction-diffusion models is closely related to integrable
magnetic chains. The relationship with the Hecke algebra and its quotients
allows to identify integrable reaction-diffusion models and, through the
Baxterization procedure, relate them to the solutions of Yang-Baxter equations
which can be solved via the Bethe ansatz. Methods such as spectral and
partial integrability, free fermions, similarity transformations or
diffusion algebras are reviewed, with several concrete
examples treated explicitly. An outlook on how the
recently-introduced concept of local scale invariance might become useful
in the description of non-equilibrium ageing phenomena is presented, with
particular emphasis on the kinetic Ising model with Glauber dynamics. 
\end{abstract}
\end{titlepage}
%%%%%%%%%%%%%%%%%%%%%%%%%%%%%%%%%%%%%%%%%%%%%%%%%%%%%%%%%%%%%%%%%%%%%%%%%%%%%%%%
%            Title page finished, we can START !!                              %
%%%%%%%%%%%%%%%%%%%%%%%%%%%%%%%%%%%%%%%%%%%%%%%%%%%%%%%%%%%%%%%%%%%%%%%%%%%%%%%%

%%%STARTSTARTSTARTSTARTSTARTSTARTSTARTSTARTSTARTSTARTSTARTSTARTSTARTSTARTSTART%%
%%%STARTSTARTSTARTSTARTSTARTSTARTSTARTSTARTSTARTSTARTSTARTSTARTSTARTSTARTSTART%%

%%%%%%%%%%%%%%%%%%%%%%%%%%%%%%%%%%%%%%%%%%%%%%%%%%%%%%%%%%%%%%%%%%%%%%%%%%%%%%%%
\section{Reaction-diffusion processes} 
%%%%%%%%%%%%%%%%%%%%%%%%%%%%%%%%%%%%%%%%%%%%%%%%%%%%%%%%%%%%%%%%%%%%%%%%%%%%%%%%

The understanding of non-equilibrium statistical physics is still much more
incomplete than that of equilibrium theory, due to the absence
of an analogue of the Boltzmann-Gibbs approach and in spite of
considerable recent progress \cite{Derr02}. Therefore,  
non-equilibrium systems have to be specified by some defining dynamical
rules which are then analyzed. The topic has received a lot of attention 
and many reviews exist, e.g. 
\cite{Derr93b,Schm95,Priv96,Marr99,Hinr00,Avra00,Cate00,Cugl02}. 
Exactly solvable systems far from equilibrium have been recently reviewed
in a nice way \cite{Schu00}. Here I present a 
pedagogically-minded introduction to the application of a few standard tools
from one-dimensional integrable quantum systems to non-equilibrium statistical
mechanics. After recalling why standard descriptions such a kinetic or
reaction-diffusion differential equations are in general insufficient in one 
dimension, we remind the reader in section~2 on the quantum Hamiltonian 
formulation of non-equilibrium processes which in turn is based on the 
master equation. Section~3 recalls a few basic facts about Hecke algebras. 
These building blocks are used in sections~4 and 5 to show explicitly the 
integrability of certain single-species reaction-diffusion processes, 
through their relation to integrable vertex models. 
Section~6 reviews some further methods such as spectral and partial 
integrability, the free-fermion technique, similarity transformations or
diffusion algebras. We close in section~7 with an outlook on how the
recently-introduced concept of local scale invariance might become useful
in the description of non-equilibrium ageing phenomena. 

I have made no effort to provide a complete bibliography. This may be
found in the excellent reviews quoted above.

%%----------------------------------------------------------------------------%%
\begin{figure}[h]
\centerline{\epsfxsize=3.05in\epsfbox
{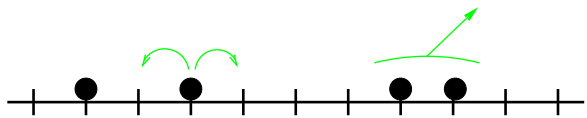}
}
\caption[Reaction model]{Microscopic reactions in the diffusive
pair-annihilation process.   
\label{Bild1}}
\end{figure}
%%----------------------------------------------------------------------------%%
A class of non-equilibrium models which are particularly simple to formulate 
are the so-called
{\em reaction-diffusion processes}. Consider the following example: particles
of a single kind (species) move on a lattice (figure~\ref{Bild1}). 
Each site of the lattice may
be either empty (denoted by $\circ$) or else be occupied by a single
particle (denoted by $\bullet$). The particles are allowed to undergo the
following movements, see figure~\ref{Bild1}, 
which involve the states of two nearest-neighbour sites:
\BEA 
\circ + \bullet \leftrightarrow \bullet + \circ & & 
\mbox{\rm diffusion, with rate $D$} 
\nonumber \\
\bullet + \bullet \rightarrow \circ + \circ & & 
\mbox{\rm pair-annihilation, with rate $2\alpha$} 
\label{1:gl:raten}
\EEA
The first of the allowed movements in (\ref{1:gl:raten}) is reversible 
while the second is not. 
A typical question is then for the long-time behaviour of quantities such as
the mean particle density $n(t)$. Trivially $n(t)$ decreases with
increasing time $t$ but different long-time asymptotic behaviours such as
$n(t)\sim t^{-\alpha}$ or $n(t)\sim e^{-t/\tau}$ are conceivable. The oldest
approach to this problem was introduced by Smoluchowski \cite{Smol17} and
consists of writing down kinetic equations, e.g. for the spatially
averaged density $n(t)$ and one obtains 
$\partial_t n(t) = -\lambda n(t)^2$ for the problem at hand, where 
$\lambda=4\alpha$. With the initial condition $n(0)=n_0$, the solution 
$n(t) = n_0 (1+n_0 \lambda t)^{-1} \simeq (\lambda t)^{-1}$
is easily found and apparently answers the physical question. A slightly
more involved version of this argument does allow for spatial variation
of the density $n=n(\vec{r},t)$ and considers a 
{\em reaction-diffusion equation}
\BEQ \label{1:gl:n}
\partial_t n(\vec{r},t) = D \vec{\nabla}^2 n(\vec{r},t) -\lambda n(\vec{r},t)^2
\EEQ
While the analysis of such non-linear partial differential equations is a 
for\-mi\-dab\-le problem in its own right, 
these equations do not yet capture the
essential physics in low-spatial dimensions, as we now show. Rather, they 
must be considered as an approximation of mean-field type. 

In order to understand the approximative nature of equations such as 
(\ref{1:gl:n}), and following \cite{Avra90}, consider the mean particle-density
in a large volume $V$ 
\BEQ
\bar{n}(t) = \frac{1}{V} \int_V \!\D\vec{r}\: n(\vec{r},t) 
\EEQ
It then follows
\BEA
\partial_t \bar{n}(t) &=& \frac{1}{V} \int_V \!\D\vec{r}\: \left[ 
D \vec{\nabla}^2 n(\vec{r},t) - \lambda n(\vec{r},t)^2 \right] 
\nonumber \\
&=& \frac{D}{V} \int_{\partial V} \!\D\vec{\sigma}\cdot 
\vec{\nabla} n(\vec{r},t)
-\frac{\lambda}{V} \int_V \!\D\vec{r}\: n(\vec{r},t)^2
\nonumber \\
&\leq& -\lambda \left( \frac{1}{V} \int_V \!\D\vec{r}\: n(\vec{r},t) \right)^2
\EEA
In the second line, Gau{\ss}' theorem was used where $\D\vec{\sigma}$ is
normal to the surface $\partial V$ (the flow through
$\partial V$ vanishes for large volumes $V\to\infty$). The
last line follows from the Cauchy-Schwarz inequality. Together with the
initial condition $\bar{n}(0)=n_0$, the inequality 
$\partial_t\bar{n}(t)\leq -\lambda \bar{n}(t)^2$ yields the
bound
\BEQ \label{1:gl:borne2}
\bar{n}(t) \leq \frac{n_0}{1+n_0 \lambda t} \leq \frac{1}{\lambda t}
\EEQ
for all times $t\geq 0$. However, the model
defined above can be solved exactly in one spatial dimension 
(in a setting defined precisely in section~3), provided $D=2\alpha$. 
The exact mean particle-density is given by \cite{Spou88}
\BEA
\bar{n}(t) &=& n_0 e^{-4Dt} \left[ I_0(4Dt) + 2(1-n_0) 
\sum_{k=1}^{\infty} (1-2n_0)^{k-1} I_k(4Dt) \right]  
\nonumber \\
&\simeq& \frac{1}{\sqrt{8\pi D\,}}\: t^{-1/2} \;\; ; \;\; t\to\infty
\label{1:gl:exakt}
\EEA
where $I_k$ is the modified Bessel function of order $k$. Clearly, for large
times the exact mean density $\bar{n}(t)$ decreases considerably slower than
even the upper bound (\ref{1:gl:borne2}) derived from the reaction-diffusion 
equation (\ref{1:gl:n}). 

The failure of eq. (\ref{1:gl:n}) to describe correctly the long-time behaviour
can be understood from the following heuristic argument \cite{Tous83}. 
In the long-time limit, the particle-density should already be very low and
it is conceivable that at most one annihilation reaction takes place at a
given time. Let $L=L(t)$ be the typical distance between two particles. Then
the time needed by diffusive motion to overcome this distance is of the
order $t\sim L(t)^2$. On the other hand, the mean particle density is
$\bar{n}(t) \sim L(t)^{-d}$ in $d$ spatial dimensions and this argument
would give $\bar{n}(t) \sim t^{-d/2}$. Therefore, the assumption implicit 
in equations such as (\ref{1:gl:n}) that diffusive motion can render the 
system sufficiently homogeneous fails in low dimensions (in our model for $d<2$)
and one rather has the long-time behaviour \cite{Tous83}
\BEQ
\bar{n}(t) \sim \left\{ \begin{array}{ll} 
t^{-d/2} & \mbox{\rm if $d<2$} \\
t^{-1}   & \mbox{\rm if $d>2$} \end{array} \right.
\EEQ
Therefore $d^*=2$ is the {\em upper critical dimension} of the diffusive
pair-anni\-hi\-la\-tion process. 
For dimensions $d>d^*$, reaction-diffusion equations
should be expected to give qualitatively correct results and entire
branches of physical chemistry are built on this. On the other hand, for
low-dimensional structures with $d<d^*$, as might occur for example in
nanodevices, fluctuation effects become dominant. 

The importance of fluctuations in low-dimensional reaction-diffusion
processes has also been confirmed experimentally. An effectively
one-dimensional setting can be achieved by studying the kinetics of
excitons (localized electronic excitations) along polymer chains (other 
examples are reviewed in \cite{Schu00,Hinr00}). For
details, consult the reviews by Kroon and Sprik and Kopelman and Lin 
in \cite{Priv96}. 
The only purpose of the polymer chains is to provide a carrier for the
excitons. Schematically, single excitons may hop from one monomer to the
next (thus modelling a diffusive motion) while a reaction occurs if two
excitons meet. One may have one or both of the reactions
$\bullet\bullet \rightarrow \circ\circ$ or 
$\bullet\bullet \rightarrow \bullet\circ$, see table \ref{tab:ReakDiffExp}. 
We shall show in section~6 that for any branching ratio 
$B=\Gamma(\bullet\bullet\rightarrow\circ\circ)/
\Gamma(\bullet\bullet\rightarrow\bullet\circ)$ the long-time behaviour is
still described by the model (\ref{1:gl:raten}), with a renormalized rate. 
For late times, one expects the mean exciton density to fall off as a
power law $\bar{n}(t) \sim t^{-y}$. The excitons are unstable, with 
lifetimes of the order $\lesssim 10^{-3} \mbox{\rm s}$. 
Their decay produces light of a characteristic frequency whose intensity can be
used to measure $\bar{n}(t)$ while light with a different frequency is
emitted if excitons decay through a pair reaction. This allows to measure
$\partial_t \bar{n}(t)$ as well through time-resolved experiments down to the
picosecond scale. Table~\ref{tab:ReakDiffExp} gives some results for the 
exponent $y$ in different materials  
(the branching ratio $B\lesssim 10\%$ in the first two lines of 
table~\ref{tab:ReakDiffExp}). 
Clearly $y\simeq 1/2$ as expected from (\ref{1:gl:exakt}) and far
from unity. This is strong evidence in favour of
strong fluctuation effects in these systems and against their description
through a reaction-diffusion equation~(\ref{1:gl:n}).

\begin{table}
\caption[Decay exponent for exciton kinetics]{Measured decay exponent $y$
of the mean exciton density $\bar{n}(t)\sim t^{-y}$ on polymer chains. 
The error bar for TMMC comes from averaging over the results of 
\protect{\cite{Kroo93}} for different initial particle densities.  
\label{tab:ReakDiffExp}}
\begin{center}
\begin{tabular}{|l|clc|} \hline
~substance & $y$ & ~reaction(s)\ & Reference \\ \hline
C$_{10}$H$_8$ & 0.52 - 0.59  & 
$\bullet\bullet\rightarrow \left\{\vekz{\circ\circ}{\bullet\circ}\;\right.$ 
& \cite{Pras89} \\
P1VN/PMMA film\,  {} & 0.47(3)  & 
$\bullet\bullet\rightarrow \left\{\vekz{\circ\circ}{\bullet\circ}\;\right.$ & 
\cite{Kope90} \\
TMMC & $0.48(4)$ & 
$\bullet\bullet\rightarrow\;\;\;\; \bullet\circ$& \cite{Kroo93} \\ \hline
\end{tabular}
\end{center}
\end{table}

Another aspect becomes apparent if we now briefly consider the triple
annihilation process $\bullet \bullet \bullet \rightarrow \circ \circ \circ$  
combined with single particle diffusion. The reaction-diffusion equation
reads
\BEQ \label{1:gl:n3}
\partial_t n(\vec{r},t) = D \vec{\nabla}^2 n(\vec{r},t) -\lambda n(\vec{r},t)^3
\EEQ
Following the same lines as above, but using now the H\"older 
inequality, the differential inequality 
$\partial_t\bar{n}(t)\leq -\lambda \bar{n}(t)^3$ leads to the bound
\BEQ
\bar{n}(t) \leq \frac{n_0}{\sqrt{1+2n_0^2 \lambda t\,}} 
\leq \frac{1}{\sqrt{2\lambda\,}}\: t^{-1/2}
\EEQ
This of the same order of magnitude as the long-time behaviour expected
from diffusive motion in one spatial dimension. Therefore, and in agreement
with scaling arguments showing that $d^*=1$ \cite{Corn93}, already for
the triple annihilation process, fluctuation effects should not play a major
role in any physically realizable dimension ($d>1$).\footnote{This does not
imply however, that models containing both binary {\em and} multisite reaction
terms could not have a non-trivial behaviour. For example, the phase structure 
of the pair contact process ($\bullet\bullet\rar\circ\circ$, 
$\bullet\bullet\circ\rar\bullet\bullet\bullet$) with single particle 
diffusion ($\bullet\circ\leftrightarrow\circ\bullet$) is presently 
controversial and under very active study \cite{Henk03a}.}

In conclusion, the description of reaction-diffusion processes with
pair reactions in low dimensions requires a truly microscopic approach beyond
kinetic reaction-diffusion equations while these equations may well turn
out to be adequate for multiparticle reactions. For that reason, we shall in
the following consider the master equation formulation of reaction-diffusion
processes with pair-reaction terms only.

%%%%%%%%%%%%%%%%%%%%%%%%%%%%%%%%%%%%%%%%%%%%%%%%%%%%%%%%%%%%%%%%%%%%%%%%%%%%%%%%
\section{Quantum Hamiltonian formulation} 
%%%%%%%%%%%%%%%%%%%%%%%%%%%%%%%%%%%%%%%%%%%%%%%%%%%%%%%%%%%%%%%%%%%%%%%%%%%%%%%%

We now review the reformulation of a non-equilibrium stochastic 
system defined by some master equation in terms 
of the spectral properties of an associated 
quantum Hamiltonian $H$ and which goes back at least to the classic
paper by Glauber \cite{Glau63}. To be specific, we consider in this section
only systems defined
on a chain with $L$ sites and two allowed states per site. We represent
the states of the system in terms of spin configurations $\vec{\sig} =
\{ \sig_1, \sig_2, \ldots , \sig_L \}$ where $\sig=+1$ corresponds to an
empty site and $\sig=-1$ corresponds to a site occupied by a single particle. 
We are interested in the probability distribution function $P(\vec{\sig};t)$
of the configurations $\vec{\sig}$. Our starting point is the master 
equation for the $P(\vec{\sig};t)$
\BEQ \label{Master}
\partial_t P(\vec{\sig};t) = \sum_{\vec{\tau}\neq\vec{\sig}} \left[
w(\vec{\tau}\rar\vec{\sig}) P(\vec{\tau};t) 
- w(\vec{\sig}\rar\vec{\tau}) P(\vec{\sig};t) 
\vekz{ }{\,} \!\!\!\!\!\right]
\EEQ
where $w(\vec{\tau}\rar\vec{\sig})$ are the transition rates between 
the configurations $\vec{\tau}$ and $\vec{\sig}$ and are assumed to be 
given from the phenomenology. 
In order to rewrite this as a matrix problem, one introduces a state vector
\BEQ
|P\rangle = \sum_{\vec{\sig}} P(\vec{\sig};t) |\vec{\sig}\rangle
\EEQ
and eq.~(\ref{Master}) becomes
\BEQ \label{2:gl:HSchr}
\partial_t \ket{P} = - H \ket{P}
\EEQ
where the matrix elements of $H$ are given by
\BEQ
\bra{\sig} H \ket{\tau} = -w(\vec{\tau}\rar\vec{\sig}) +\delta_{\sig,\tau}
\sum_{\vec{\ups}} w(\vec{\tau}\rar\vec{\ups})
\EEQ
The operator $H$ describes a stochastic process since all the
elements of the columns add up to zero. Conservation of probability
$\sum_{\vec{\sig}} P(\vec{\sig};t) =1$ is equivalent to the relation
\BEQ
\bra{s} H = 0
\EEQ
where $\bra{s} = \sum_{\vec{\sig}} \bra{\vec{\sig}}$ is a {\em left} 
eigenvector of $H$ with eigenvalue 0. Correspondingly, 
$H$ has at least one right eigenvector 
$\ket{s}=\sum_{\vec{\sig}} P_s(\vec{\sig})\ket{\vec{\sig}}$ with 
eigenvalue 0, that is $H\ket{s}=0$. 
Such a vector does not evolve in time and therefore corresponds 
to a steady-state distribution of the system. Since in general $H$ is {\em not} 
symmetric, this steady-state vector may be highly non-trivial. Note that all
this is completely general and applies to any stochastic process defined by
a master equation. With a view on the processes to be studied below one calls
$H$ a {\em quantum Hamiltonian} and this formulation of the master equation
the {\em quantum Hamiltonian formalism} (see \cite{Hinr00,Schu00} for recent
reviews). The reason for this choice of language
is the fact that for the processes studied below (and, in fact, many other
processes as well) $H$ is the Hamiltonian of some quantum system such as
the Heisenberg XXZ Hamiltonian. The steady-state $\ket{s}$ of a stochastic
system corresponds in this mapping to the ground state of the quantum system. 
The probabilistic interpretation is guaranteed by the following classical
result.\\

\noindent
{\bf Theorem 1} (Hyver, Keizer, Schnackenberg \cite{Hyver}): 
{\it For a quantum Hamiltonian $H$ which satisfies the
master equation (\ref{2:gl:HSchr}) and has 
$\bra{s}=\sum_{\vec{\sig}}\bra{\vec{\sig}}$ as a left eigenstate
such that $\bra{s} H =0$, the following holds. (i) There is a stationary
state}
\BEQ
\ket{s} = \sum_{\vec{\sig}} P_s(\vec{\sig}) \ket{\vec{\sig}}
\EEQ
{\it such that $H\ket{s}=0$. (ii) Consider the eigenvalue problem
$H\ket{n} = E_n \ket{n}$, with $n=0,1,2\ldots$. Then}
\BEQ \label{2:gl:Th1-2}
\Re E_n \geq E_0 =0
\EEQ
{\it (iii) Let $\ket{P_0} = \ket{P(0)}$ be the initial state such that
the weights of the individual configurations satisfy 
$0\leq P(\vec{\sig};0)\leq 1$
and $\braket{s}{P_0}=1$. Then for all times $t\geq 0$, one has}
\BEQ
0\leq P(\vec{\sig};t) \leq 1 \;\; \mbox{\rm ~~ and~~} \braket{s}{P}=1
\EEQ
{\it (iv) Let $H: \mathbb{R}^n\to\mathbb{R}^n$ be a linear map such that for 
the elements $H_{ij}$ of $H$ holds}
\BEQ \label{2:gl:StoBed}
H_{ij} \leq 0 \;\; , \;\; \sum_{i=1}^{n} H_{ij} = 0 ~~~ 
\forall j\in\{1,\ldots,n\}
\EEQ
{\it Then $H$ is a ``quantum Hamiltonian'' of a Markov process described by 
the master equation (\ref{2:gl:HSchr}).}\\

\noindent 
Time-dependent averages of an observable $F$ are given by 
the matrix element
\BEQ
<F>(t) = \sum_{\vec{\sig}} F(\vec{\sig}) P(\vec{\sig};t)  
= \bra{s} F \ket{P} = \bra{s} F \exp(-H t)\ket{P_0}
\label{AVE}
\EEQ
and we see that eq.~(\ref{2:gl:Th1-2}) means that the system indeed evolves 
towards the steady-state $\ket{s}$, 
thus $P_s(\vec{\sig})=P(\vec{\sig};\infty)$.

In what follows, we shall be mainly 
interested in averages of particle numbers
$n_{j}$ at site $j$ and their correlators. These can be expressed in the
quantum spin formulation in terms of the projector
\BEQ \label{NOp}
\tilde{n}_{j} = \frac{1}{2} \left( 1 - \sig_j^z \right) = 
\left( \matz{0}{0}{0}{1} \right)_j
\EEQ
and one-point and two-point functions of the $n_j$ are then 
expressed as\footnote{We stress that the structure of these matrix 
elements is quite distinct from expectation values $\bra{0} F \ket{0}$ in 
ordinary quantum mechanics.}
\BEQ \label{OneTwo}
C_1(j;t) =\, <\!\tilde{n}_j\!>(t) \,= \bra{s} \tilde{n}_j \ket{P} \;\; , \;\; 
C_2(j,\ell;t) =\, <\!\tilde{n}_j \tilde{n}_{\ell}\!>(t) \,= \bra{s}\tilde{n}_j 
\tilde{n}_{\ell} \ket{P}
\EEQ 

Two basic situations are readily distinguished from the spectrum of $H$. If
in the limit of infinite lattice size $L\rar\infty$ the lowest excited states
have a finite gap $\Gamma$ to the ground state energy $E_0=0$, 
then the averages
(\ref{OneTwo}) will approach their steady state values exponentially fast
on the time-scale $\tau=1/\Gamma$. On the other hand, if there is in the 
$L\rar\infty$ limit a continuous spectrum down to $E_0=0$, one expects an 
algebraic decay of the correlators as the system approaches the steady state.  

%%%%%%%%%%%%%%%%%%%%%%%%%%%%%%%%%%%%%%%%%%%%%%%%%%%%%%%%%%%%%%%%%%%%%%%%%%%%%%%%
\section{Hecke algebra and integrability}
%%%%%%%%%%%%%%%%%%%%%%%%%%%%%%%%%%%%%%%%%%%%%%%%%%%%%%%%%%%%%%%%%%%%%%%%%%%%%%%%

Before we shall write down quantum Hamiltonians for certain 
reac\-tion-dif\-fu\-sion
systems explicitly, we need some background information on Hecke algebras in
order to make contact with integrability. The Hecke algebra $H_n(q)$ is
spanned by $n$ generators $e_i$
\BEQ
H_n(q) = \left\{ e_1, e_2, \ldots, e_n \right\}
\EEQ
which satisfy the following relations
\BEA
e_i e_{i\pm 1} e_i - e_i &=& e_{i\pm 1} e_i e_{i\pm 1} - e_{i\pm 1} 
\nonumber \\
e_i e_j &=& e_j e_i \;\; ; \;\; \mbox{\rm ~~if $|i-j|\geq 2$}
\label{3:gl:Hecke} \\
e_i^2 &=& \left( q + q^{-1}\right) e_i 
\nonumber 
\EEA
where $q\in\mathbb{C}$ is a parameter. The representations of $H_n(q)$ and the
relationship to equilibrium statistical mechanics are discussed in great detail
in \cite{Mart91}. We are not only interested in $H_n(q)$, but also in some
{\em quotients}, denoted by $(P,M)H_n(q)$ \cite{Mart92}. Two specific examples
will be of interest to us. The first such quotient is the 
celebrated Temperley-Lieb algebra $(2,0)H_n(q)$, where in addition to
eq.~(\ref{3:gl:Hecke}) the additional relations
\BEQ
e_i e_{i\pm 1} e_i - e_i =0 \;\; , \;\; 
e_{i\pm 1} e_i e_{i\pm 1} - e_{i\pm 1} = 0
\EEQ
hold. On the other hand, the quotient $(1,1)H_n(q)$ is defined through the
condition \cite{Mart92}
\BEQ 
\left( e_{i}e_{i+2}\right) e_{i+1} \left(q+q^{-1}-e_{i}\right)
\left(q+q^{-1}-e_{i+2}\right) =0 \;\; ; \;\; i=1,2,\ldots
\EEQ
For the definition of more general quotients, we refer to \cite{Mart92}. 

Consider the $N\times N$ matrices $E^{ab}$ where $a,b=0,1,\ldots,N-1$. The 
only non-vanishing element of $E^{ab}$ is the one on the $a^{\rm th}$ line
and the $b^{\rm th}$ column and this element is equal to unity. Define
further
\BEQ \label{3:gl:Eiab}
E_i^{ab} = {\bf 1} \otimes \cdots \otimes {\bf 1} \otimes
E^{ab} \otimes {\bf 1} \otimes \cdots \otimes {\bf 1}
\EEQ
where $E^{ab}$ occurs on position $i$ and $i=1,\ldots,L$ runs over the
sites of a chain. Explicit realizations of the Hecke algebra or one of
its quotients may be found in the Perk-Shultz models \cite{Shul81}, 
whose Hamiltonian is of the form
\BEQ \label{3:gl:PS}
H^{(P,M)} = \sum_{i=1}^{L-1} \, e_i^{(P,M)} 
\EEQ 
The importance of this observation becomes clear from the following \\

\noindent
{\bf Theorem 2} (Jones \cite{Jone89}): 
{\it If $H=\sum_{i=1}^{L-1} e_i$, where the $e_i$ are the generators of the
Hecke algebra $H_{L-1}(q)$, $H$ is integrable through the Baxterization
procedure.} \\

\noindent The Baxterization procedure allows to define, starting from $H$, 
in a systematic way Boltzmann weights which satisfy the Yang-Baxter equation. 
We shall illustrate this in the example of the seven-vertex model in
section~5. 

In many practical applications, the following result is useful.\\ 

\noindent
{\bf Theorem 3} (Martin, Rittenberg \cite{Mart92}): 
{\it If $H=\sum_{i=1}^{L-1} e_i$, where the $e_i$ are the generators of the
Hecke algebra quotient $(P,M) H_{L-1}(q)$ and furthermore 
$H'=\sum_{i=1}^{L-1} f_i$, where the $f_i$ are different generators of the
same quotient $(P,M) H_{L-1}(q)$, then $H$ and $H'$ have the same eigenvalues,
up to degeneracies.} \\ 

\noindent 
We finish this section by writing explicit examples for the quotients
$(2,0)$ and $(1,1)$ in the case $N=2$ \cite{Alca93}. Then the matrices 
$E^{ab}$ can be expressed through Pauli matrices
\BEQ \label{3:gl:Pauli}
\sig^x = \left( \matz{0}{1}{1}{0} \right) \;\; , \;\;
\sig^y = \left( \matz{0}{-\II}{\II}{0} \right) \;\; , \;\;
\sig^z = \left( \matz{1}{0}{0}{-1} \right)
\EEQ
and we define $\sig_i^{x,y,z}$ by analogy with (\ref{3:gl:Eiab}). Set
\BEQ \label{3:gl:e20}
e_i^{(2,0)} = -\frac{1}{2} \left[ \sig_i^x \sig_{i+1}^x +
\sig_i^y \sig_{i+1}^y +\frac{q+q^{-1}}{2} \left( 
\sig_i^z \sig_{i+1}^z - 1\right) - \frac{q-q^{-1}}{2}
\left( \sig_i^z -\sig_{i+1}^z\right) \right] 
\EEQ
Therefore, the Hamiltonian \cite{Kiri88,Pasq90}
\BEQ \label{3:gl:H20}
H^{(2,0)} = \sum_{i=1}^{L-1} e_i^{(2,0)} 
\EEQ
is integrable. In addition, it satisfies a quantum group invariance, since
\BEQ
\left[ H^{(2,0)}, S^z \right] = \left[ H^{(2,0)}, S^{\pm} \right] = 0
\EEQ
where, recalling also that $\sig^{\pm}=\frac{1}{2}(\sig^x\pm \II\sig^y)$
\BEA
\lefteqn{ S^z = \frac{1}{2} \sum_{i=1}^{L} \sig_i^z \;\; , \;\;
S^{\pm} = \sum_{i=1}^{L} S_i^{\pm} }
\nonumber \\
S_i^{\pm} &=& \exp\left(\frac{\ln q}{2}\sum_{\ell=0}^{i-1}\sig_{\ell}^z\right)
\sig_i^{\pm}  
\exp\left(-\frac{\ln q}{2} \sum_{\ell=i+1}^{L} \sig_{\ell}^z\right)
\EEA
which in turn obey the commutation relations of $U_q(su(2))$, namely
\BEQ
\left[ S^z, S^{\pm} \right] = \pm S^{\pm} \;\; , \;\;
\left[ S^+, S^- \right] = \frac{q^{2S^z} - q^{-2S^z}}{q-q^{-1}} 
\EEQ
On the other hand, set
\BEQ \label{3:gl:e11}
e_i^{(1,1)} = -\frac{1}{2} \left[ \sig_i^x \sig_{i+1}^x +
\sig_i^y \sig_{i+1}^y +
q \sig_i^z + q^{-1} \sig_{i+1}^z -q -q^{-1} \right] 
\EEQ
Here, the associated integrable Hamiltonian \cite{Sale89} 
$H^{(1,1)}=\sum_{i=1}^{L-1} e_i^{(1,1)}$ in addition is invariant under the
supersymmetric quantum group $U_q(su(1|1))$, since
\BEQ
\left[ H^{(1,1)}, S^z \right] = \left[ H^{(1,1)}, T^{\pm} \right] = 0
\EEQ
where
\BEQ
T^{\pm} = q^{(1-L)/2} \sum_{i=1}^{L} q^{i-1} \exp\left(\frac{\II\pi}{2}
\sum_{\ell=1}^{i-1} (\sig_{\ell}^z+1)\right) \sig_i^{\pm}
\EEQ
and
\BEQ
\left[ S^z, T^{\pm} \right] = \pm T^{\pm} \;\; , \;\;
\left\{ T^+, T^- \right\} = \frac{q^{L} - q^{-L}}{q-q^{-1}} 
\EEQ

%%%%%%%%%%%%%%%%%%%%%%%%%%%%%%%%%%%%%%%%%%%%%%%%%%%%%%%%%%%%%%%%%%%%%%%%%%%%%%%%
\section{Single-species models} 
%%%%%%%%%%%%%%%%%%%%%%%%%%%%%%%%%%%%%%%%%%%%%%%%%%%%%%%%%%%%%%%%%%%%%%%%%%%%%%%%
  
We are now ready to study explicit examples of stochastic quantum Hamiltonians. 
The classical example merely considers particles of a single species ($\bullet$)
which may hop randomly onto an empty nearest-neighbour site ($\circ$), thereby 
modelling the reversible reaction $\bullet\circ\leftrightarrow\circ\bullet$
with rate $D$. This process is often called the {\em symmetric exclusion
process}. The quantum Hamiltonian reads
\BEQ
H = -\frac{D}{2} \sum_{j=1}^{L-1} \left[ \sig_j^x \sig_{j+1}^x 
+\sig_j^y \sig_{j+1}^y + \left( \sig_j^z\sig_{j+1}^z -1 \right) \right]
\EEQ
and co\"{\i}ncides with the (ferromagnetic) XXX Heisenberg quantum 
chain \cite{Alex78}. 

Certainly, one may now use the Bethe ansatz solution of $H_{\rm XXX}$
to rederive known results on simple diffusion. 
The recent interest in this setup comes from the insight that the 
{\em integrability} of the associated quantum chains allow to make contact
with the pre-established algebraic techniques for the treatment of these
\cite{Alca93,Alca94}. Independently, integrability was also observed to occur
in the transfer matrices for discrete-time dynamics \cite{Kand90,Schu93}. 
The enormous possibilities for non-trivial applications then triggered an
ongoing wave of activity, see e.g. 
\cite{Derr93b,Priv96,Schu00,Alca93,Alca94a,Schu95,Albe98,Alca98,Gros03,Isae01} 
and references therein. 

\begin{table}
\begin{center}
\begin{tabular}{|l|c|c|c|c|c|c|} \hline
{\bf d}iffusion to the left & $\circ\bullet\rar\bullet\circ$ & $D_L$ 
&$a_{32}$ & $w_{32}$ & $w_{1,1}(1,0)$ & $\Gamma_{10}^{01}$ \\
{\bf d}iffusion to the right& $\bullet\circ\rar\circ\bullet$  & $D_R$
&$a_{23}$ & $w_{23}$ & $w_{1,1}(0,1)$ & $\Gamma_{01}^{10}$ \\
pair {\bf a}nnihilation     & $\bullet\bullet\rar\circ\circ$   & $2\alpha$ 
&$a_{14}$ & $w_{14}$ & $w_{1,1}(0,0)$ & $\Gamma_{00}^{11}$ \\
{\bf c}oagulation to the right& $\bullet\bullet\rar\circ\bullet$ &$\gamma_R$
&$a_{24}$ & $w_{24}$ & $w_{1,0}(0,1)$ & $\Gamma_{01}^{11}$ \\
{\bf c}oagulation to the left & $\bullet\bullet\rar\bullet\circ$&$\gamma_L$
&$a_{34}$ & $w_{34}$ & $w_{0,1}(1,0)$ & $\Gamma_{10}^{11}$ \\
{\bf d}eath at the left  & $\bullet\circ\rar\circ\circ$ & $\delta_L$
&$a_{13}$ & $w_{13}$ & $w_{1,0}(0,0)$ & $\Gamma_{00}^{10}$ \\
{\bf d}eath at the right & $\circ\bullet\rar\circ\circ$&$\delta_R$
&$a_{12}$ & $w_{12}$ & $w_{0,1}(0,0)$ & $\Gamma_{00}^{01}$ \\
decoagulation to the left & $\circ\bullet\rar\bullet\bullet$&$\beta_L$&$a_{42}$
& $w_{42}$ & $w_{1,0}(1,1)$ & $\Gamma_{11}^{01}$ \\
decoagulation to the right & $\bullet\circ\rar\bullet\bullet$&$\beta_R$&$a_{43}$
& $w_{43}$ & $w_{0,1}(1,1)$ & $\Gamma_{11}^{10}$ \\
birth at the right&$\circ\circ\rar\circ\bullet$ & $\nu_R$&$a_{21}$
& $w_{21}$ & $w_{0,1}(0,1)$ & $\Gamma_{01}^{00}$ \\
birth at the left&$\circ\circ\rar \bullet\circ$ & $\nu_L$&$a_{31}$
& $w_{31}$ & $w_{1,0}(1,0)$ & $\Gamma_{10}^{00}$ \\ 
pair creation        & $\circ\circ\rar \bullet\bullet$ & $2\sig$ &$a_{41}$
& $w_{41}$ & $w_{1,1}(1,1)$ & $\Gamma_{11}^{00}$ \\ \hline
\multicolumn{2}{|l|}{Rates defined after reference} & \cite{Henk97} &
\cite{Schu95} & \cite{Schu00} & \cite{Alca94} & \cite{Pesc94} \\ \hline
\end{tabular}
\caption{Two-sites reaction-diffusion processes of a single species 
and their rates as denoted by various authors. \label{tab1}}
\end{center} \end{table}

Following \cite{Alca93,Alca94}, we now give more examples of integrable 
quantum Hamiltonians of stochastic
systems, restricting ourselves for simplicity to a single species of
particles and to binary reactions only (see section 1). The reaction
rates are defined in table~\ref{tab1}, using the convention of various
authors, but unfortunately there is no standard notation. 
While I prefer a light notation (slightly modified from \cite{Henk97})
and shall use it here,\footnote{The letter $\nu$ is inspired by 
{\it {\bf n}aissance} (French for birth) and $\sigma$ comes from
{\it{\bf S}ch\"opfung} (German for creation). The letter $\beta$ might have
come from {\bf b}ranching.} other authors often opt for a 
systematic, though heavier notation with several indices. 

For the time being and for purposes of illustration let us consider besides 
diffusion only those reactions which irreversibly reduce
the number of particles (that is, $\beta_{L,R}=\nu_{L,R}=\sig=0$). Define
\BEQ 
D = \sqrt{ D_L D_R } \;\; , \;\; \gamma = \frac{ \sqrt{\gamma_L\gamma_R} }{D}
\;\; , \;\; \delta = \frac{ \sqrt{\delta_L\delta_R} }{D} \;\; , \;\;
q = \sqrt{ \frac{D_L}{D_R} } = \sqrt{ \frac{\gamma_L}{\gamma_R} } =
\sqrt{ \frac{\delta_L}{\delta_R} } 
\EEQ
\BEQ \label{hDelta}
\Delta = \frac{1}{2} \left( q+q^{-1}\right)(1+\delta-\gamma) - \alpha/D 
\;\; , \;\;
h = \frac{1}{2}\left( 2\alpha/D + \gamma\left(q+q^{-1}\right) \right) .
\EEQ
Note that the ratio of the left and right rates is taken to be the same for
diffusion, coagulation and death processes. We first consider an open chain
with $L$ sites. Then the quantum Hamiltonian becomes
\BEQ \label{4:gl:HDacd}
H =  D \left( \, H_{\rm XXZ}(h,\Delta,\delta) + 
H_{\alpha} + H_{\gamma} + H_{\delta}\vekz{ }{ }\!\!\! \right)
\EEQ
where $H_{\rm XXZ}(h,\Delta,\delta)$ is the standard XXZ quantum chain,
including bulk and boundary magnetic fields
\BEA
H_{\rm XXZ}(h,\Delta,\delta) &=& -\frac{1}{2} \sum_{j=1}^{L-1} 
\left[ \sig_j^x \sig_{j+1}^x 
+\sig_j^y \sig_{j+1}^y + \Delta \left( \sig_j^z\sig_{j+1}^z -1 \right)
\right. 
\\
&+& h \left. \left( \sig_j^z + \sig_{j+1}^z -2 \right)
- \frac{1}{2} (1-\delta) \left( q-q^{-1} \right) 
\left( \sig_j^z - \sig_{j+1}^z \right) \right]
\nonumber 
\EEA
which contains the diagonal and diffusive matrix elements while the particle
annihilation terms are contained in 
\BEA
H_{\alpha} &=& -2\alpha \sum_{j=1}^{L-1} q^{-2j-1} \sig_{j}^{+} \sig_{j+1}^{+}
\nonumber \\
H_{\gamma} &=& -\gamma \sum_{j=1}^{L-1} q^{-j}
\left( \tilde{n}_j\sig_{j+1}^+ + q^{-1}\sig_{j}^+ \tilde{n}_{j+1}\right) 
\label{4:gl:ParReac} \\
H_{\delta} &=& -\delta \sum_{j=1}^{L-1} q^{-j} 
\left( q^{-2}(1-\tilde{n}_j)\sig_{j+1}^+ + 
q\sig_{j}^+ (1-\tilde{n}_{j+1})\right)
\nonumber 
\EEA
and $\sig^{\pm}=\frac{1}{2}(\sig^x\pm \II\sig^y)$ are the one-particle 
annihilation/creation operators. 

For a physical understanding of this we consider two special cases. 
\begin{enumerate}
\item Consider pure asymmetric diffusion, that is $\alpha=\gamma=\delta=0$,
also referred to as the {\em asymmetric exclusion process}. 
Then $H=D H^{(2,0)}$ as given by eqs.~(\ref{3:gl:e20},\ref{3:gl:H20}). We thus
have a very clear physical interpretation of the quantum-group parameter
$q=\sqrt{D_R/D_L}$ \cite{Alca94}. Besides simple biased diffusion, 
this model is related e.g. to the $1D$ Kardar-Parisi-Zhang equation 
or to the noisy Burgers equation \cite{Gwa92}. 
The quantum group may be used for the calculation of 
correlation functions \cite{Sand94}.
\item In addition to diffusion, add annihilation such that $2\alpha=D_L+D_R$,
that is $\Delta=0$ and keep $\gamma=\delta=0$. Then $H=D (H_0+ H_1)$, where
the hermitian part $H_0 = H^{(1,1)}$ is given by eq.~(\ref{3:gl:e11})
and this part alone is therefore supersymmetric. On the other hand, 
$H=D\sum_{i=1}^{L-1} f_i$, where $f_i \in (1,1)H_{L-1}(q)$, but the $f_i$
are no longer symmetric. This was the first example of a non-symmetric 
realization of a Hecke algebra \cite{Alca94}. We remark that besides the 
already established integrability, this system is also soluble through 
free-fermion techniques.
\end{enumerate} 

\noindent
{\bf Proposition 1} \cite{Alca94}: {\it The spectrum of $H$ 
in eq.~(\ref{4:gl:HDacd}) 
is independent of the particle-reaction terms contained in 
eq.~(\ref{4:gl:ParReac}), that is}
\BEQ
\mbox{\rm spec}(H) = \mbox{\rm spec}(D H_{\rm XXZ}(h,\Delta,\delta))
\EEQ

\noindent 
To see this, recall that the XXZ Hamiltonian conserves the number of particles 
while the reaction terms irreversibly decrease the total particle number. 
Thus, $H$ can be written in a block diagonal form
\BEQ
H = \left( \begin{array}{ccccc} 
{\cal N}_0 & X_{\delta} & X_{\alpha} & & \\
 & {\cal N}_1 & X_{\gamma,\delta} & X_{\alpha} &\\
 & & {\cal N}_2 & X_{\gamma,\delta}& \ddots \\
 & & & \ddots & \ddots\end{array} \right)
\EEQ
where ${\cal N}_n$ refers to the $n$-particle states and 
$X$ are the reaction matrix
elements. Because of the identity
\BEQ
\det \left( \begin{array}{cc} {\cal A} & X \\ 0 & {\cal B} \end{array}\right) 
= \det {\cal A} \det {\cal B}
\EEQ
it follows that the elements of (\ref{4:gl:ParReac}) do not enter into the 
characteristic polynomial of $H$. \hfill q.e.d.

Therefore, the phase diagram for the full Hamiltonian $H$ can be read off from 
the well-known spectrum of $H_{\rm XXZ}(h,\Delta,\delta)$ \cite{John72}. 
For our purposes, we need the following \cite{Alca94}. From (\ref{hDelta}), 
only the portion of the phase diagram where $h+\Delta\geq 1$ is important 
for us. First, he spectrum always has a finite gap when $h+\Delta >1$, 
which is realized whenever $\delta\neq 0$ or $q\neq 1$. Then the ground
state of $H_{\rm XXZ}$ is a trivial ferromagnetic frozen state which
corresponds to the empty state $\circ\circ\ldots\circ\circ$. 
The energy gap $\Gamma=E_1-E_0$ is finite. 
Second, the spectrum is gapless for $\Delta+h=1$,
where the system undergoes a Pokrovsky-Talapov transition. This situation
occurs for $\delta=0$ and $q=1$. We have thus identified the
cases where the model approaches the steady state exponentially
(non-vanishing gap) or algebraically (gapless). 

At this point, it is of interest to discuss the r\^ole of the boundary
conditions and we consider now a {\em periodic chain}, for simplicity just for
the asymmetric exclusion process (that is $\alpha=\gamma=\delta=0$).  
We stress that if $q\ne 1$, $H_{\rm per}$ cannot be read from 
(\ref{4:gl:HDacd}) by simply taking periodic boundary conditions. Rather,
we have
\BEQ
H_{\rm per} = - \frac{1}{q+q^{-1}} 
\sum_{i=1}^{L} \left[ q \sig_i^{+}\sig_{i+1}^{-} +
q^{-1} \sig_i^{-}\sig_{i+1}^{+} 
+ \frac{q+q^{-1}}{4} \left( \sig_i^z\sig_{i+1}^z -1 \right) \right]
\EEQ
together with the periodic boundary conditions $\sig_{L+1}^{\pm}=\sig_1^{\pm}$
and $\sig_{L+1}^z = \sig_1^z$. The hopping terms can be brought back to the 
familiar XXZ form of eq.~(\ref{4:gl:HDacd}) through a similarity transformation
$H_{\rm per}' = {\cal U} H_{\rm per} {\cal U}^{-1}$ 
with the matrix 
${\cal U}=\exp\left(\pi g\sum_{\ell=1}^{L}\ell\sig_{\ell}^z\right)$
with $q=e^{2\pi g}$ such that \cite{Henk94b}
\BEQ
H_{\rm per}'  = - \frac{1}{2(q+q^{-1})} 
\sum_{i=1}^{L} \left[\sig_i^{x}\sig_{i+1}^{x} +\sig_i^{y}\sig_{i+1}^{y} 
+ \frac{q+q^{-1}}{2} \left( \sig_i^z\sig_{i+1}^z -1 \right) \right]
\EEQ
which looks the same as $D H^{(2,0)}$, but we now have the non-periodic 
boundary conditions $\sig_{L+1}^{\pm} = q^{\mp L}\sig_1^{\pm}$, 
$\sig_{L+1}^z = \sig_1^z$. As a consequence, 
$\mbox{\rm spec}(H_{\rm per})$ has
{\em no} gap even for $q\ne 1$. While for a finite number $r$
of particles and long chains $L\to\infty$, it is easy to see that 
$\Re E_{\rm per}\sim L^{-2}$ \cite{Henk94b}, for finite densities $n=r/L$, 
an elaborate Bethe ansatz calculation shows that 
$\Re E_{\rm per}\sim L^{-3/2}$ \cite{Gwa92}.
Therefore, and quite in distinction with equilibrium systems, a change
in the boundary conditions may well induce a phase transition in the long-time
behaviour (observed first in driven diffusive systems \cite{Krug91}). 
What happens is easily understood in this particular example. For
an open chain, the particles get stuck at one end of the chain and a 
non-trivial position-dependent steady-state density-profile $\rho_s(i)$ builds 
up. The time needed for this should be of the order of the time the particles 
need to move from one end to the other which is finite for $q\ne 1$. 
On the other hand, for a periodic chain the particles keep chasing other 
forever and a steady-state particle current will be observed. By going to a 
reference frame co-moving with the mean velocity of that current, 
one is back to the case $q=1$ of unbiased diffusion.

The energy gaps $\Gamma$ can now be found by concentrating on the 
spectrum-generating part $H_{\rm XXZ}$. For $h+\Delta>1$, the gaps are finite
and are easily found. We concentrate here on the case of unbiased diffusion, 
when $h+\Delta=1$ and we are at the Pokrovsky-Talapov transition. The low-lying
energy gaps are given by the following

\noindent
{\bf Proposition 2}: {\it On the Pokrovsky-Talapov line 
$h+\Delta=1$, $\delta=0$, 
and for $L$ large, the low-lying eigenvalues of $H$ eq.~(\ref{4:gl:HDacd}) are 
for periodic boundary conditions}
\BEQ \label{4:gl:Eper}
E_r^{\rm (per)} = 
D \left( \frac{2\pi}{L}\right)^2 \left( I_1^2 + \ldots + I_{r}^2\right)
-\frac{8\pi^3 D}{L^3} \frac{\Delta}{1-\Delta} \sum_{j,\ell=1}^{r} 
\left( I_j - I_{\ell} \right)^2 + O\left( L^{-4}\right)
\EEQ
{\it where the $I_j$ are pairwise distinct integers (half-integers) when $r$ is
odd (even). For an open chain}
\BEQ \label{4:gl:Efree}
E_r^{\rm (free)} = D \left( \frac{\pi}{L}\right)^2 
\left( I_1^2 + \ldots + I_{r}^2\right) \cdot 
\left( 1 - \frac{2(r-1)}{L}\frac{\Delta}{1-\Delta} \right) 
+ O\left( L^{-4}\right)
\EEQ
{\it where the $I_j$ are pairwise distinct non-negative integers. The
integer $r=0,1,2,\ldots$ gives the number of particles in the sectors of
$H_{\rm XXZ}$.}

\noindent 
The finite-size amplitudes $\lim_{L\to\infty} L^2 E_r$
are independent of $\Delta$, for either periodic or open chains. 
This proves an old conjecture \cite{Alca94} based
on numerical calculations. For $\Delta=0$, the well-known free-fermion solution
is reproduced. To leading order in $L^{-1}$, eigenvalues with the same 
$I_1^2+\ldots+I_r^2$ are degenerate. We observe that for periodic boundary 
conditions, this degeneracy is already broken by the first correction in $1/L$, 
while for free boundary conditions, the leading correction keeps that symmetry.

Eqs.~(\ref{4:gl:Eper},\ref{4:gl:Efree}) are easily found from the Bethe ansatz. 
We first consider periodic boundary conditions. The XXZ chain may broken into
sectors containing only the states with $r$ particles. 
Performing the Bethe ansatz as usual \cite{Alca88}, one has for the
energies
\BEQ \label{4:gl:Er}
E_r = 2D \left( r - \cos k_1 -\ldots - \cos k_r\right)
\EEQ
where the quasimomenta $k_1,\ldots,k_r$ are solutions of the Bethe ansatz
equations
\BEQ
L k_j = 2\pi I_j - \sum_{\ell=1}^{r} \Theta(k_j, k_{\ell}) \;\; ; \;\;
j=1,\ldots,r
\EEQ
where the $I_j$ are pairwise distinct integers (half-integers) when $r$ is
odd (even) and
\BEQ
\Theta(k,k') = 2 \arctan \frac{\Delta \sin( (k-k')/2 )}
{\cos((k+k')/2)-\Delta\cos((k-k')/2)}
\EEQ
We are interested in the leading finite-size corrections when $L\to\infty$
with $r$ fixed. The ansatz 
\BEQ \label{4:gl:kj-ansatz}
k_j = \frac{2\pi}{L} I_j + \frac{a_j}{L^2} + \ldots
\EEQ
gives $\Theta(k_1,k_2)\simeq \frac{\Delta}{1-\Delta}\frac{2\pi}{L}
(I_1-I_2) + O(L^{-2})$ and 
$a_j =\frac{2\pi\Delta}{1-\Delta}\sum_{\ell=1}^{r} (I_{\ell}-I_{j})$.
Then (\ref{4:gl:Eper}) follows. 
Second, for free boundary conditions, the Bethe ansatz \cite{Alca87} reproduces 
eq.~(\ref{4:gl:Er}) for the energies, while the Bethe ansatz equations for
the quasimomenta now take the form
\BEQ
L k_j = \pi I_j - \frac{1}{2} \sum_{\ell\ne j} \left[ 
\Theta( k_j, k_{\ell}) - \Theta(-k_j, k_{\ell}) \right] 
\EEQ
for $j=1,\ldots,r$ and where the $I_j$ are pairwise distinct 
non-negative integers. Eq.~(\ref{4:gl:kj-ansatz}) leads to
$a_j=-\frac{\Delta}{1-\Delta}(r-1)\pi I_j^2$ and we arrive 
at (\ref{4:gl:Efree}). \hfill q.e.d. 

%%%%%%%%%%%%%%%%%%%%%%%%%%%%%%%%%%%%%%%%%%%%%%%%%%%%%%%%%%%%%%%%%%%%%%%%%%%%%%%%
\section{The seven-vertex model} 
%%%%%%%%%%%%%%%%%%%%%%%%%%%%%%%%%%%%%%%%%%%%%%%%%%%%%%%%%%%%%%%%%%%%%%%%%%%%%%%%

Having seen for some single-species reaction-diffusion processes how
the relationship with the Bethe-ansatz solution XXZ chain could be used 
to infer certain physical properties, we present in this section an
example of the Baxterization procedure \cite{Jone89}. That procedure
permits to associate to a stochastic quantum Hamiltonian $H$
related to a Hecke algebra the Boltzmann weights of a
corresponding two-dimensional vertex model and thus prove its integrability.
In principle, the model is then solved through the Bethe ansatz. For the 
six- and eight-vertex models the completeness of the Bethe ansatz has recently 
been proven \cite{Baxt02}. 

Following \cite{Alca94}, we consider the pair-annihilation model 
eq.~(\ref{4:gl:HDacd}) already defined in section~4 with $\gamma=\delta=0$
and furthermore, we take $2\alpha=D_R+D_L$, thus $\Delta=0$. We call
$\Omega=2\alpha/D$. 
After having performed the canonical transformation
$E_{i}^{ab} \rar (-1)^{a-b} E_{i}^{ab}$, only at even sites $i$,
the quantum Hamiltonian takes in the basis given by eq.~(\ref{3:gl:Pauli}) 
the simple form
\BEQ \label{eq:B4}
H = - D \sum_{i=1}^{L-1} e_i
\EEQ
where
\BEQ \label{5:gl:ei}
e_i = {\bf 1}_{1} \otimes \ldots \otimes {\bf 1}_{i-1} \otimes
\left( \begin{array}{cccc}
0 & 0 & 0 & \Omega \\
0 & q^{-1} & q & 0 \\
0 & q^{-1} & q & 0 \\
0 & 0 & 0 & q + q^{-1} \end{array}
\right) \otimes {\bf 1}_{i+2} \otimes \ldots
\EEQ
and ${\bf 1}_{i}$ are $2\times 2$ unit matrices attached to the site $i$. 

While we could certainly solve this particular model through a Jordan-Wigner
transformation followed by a canonical transformation, see e.g. \cite{Henk97},
we are interested in generic approaches of a more general validity than in 
those cases reducible to a free-fermion description. 
 
We have already seen in section~4 that the $e_i$ satisfy the Hecke algebra
(\ref{3:gl:Hecke}). We now construct a two-dimensional vertex model 
corresponding to $H$ having a row-to-row transfer matrix $T(\theta)$ 
depending on the spectral para\-me\-ter 
$\theta$. That transfer matrices will satisfy the Yang-Baxter 
equations \cite{Baxt82} which imply the commutation relations
$[T(\theta),T(\theta')]=0$ if $\theta\ne\theta'$. The construction is based 
on the matrix $\vm{R}_{i} (\theta)$, $i=1,2,\ldots,L-1$ which depends on the 
spectral parameter $\theta$. The Baxterization procedure for Hecke
algebras \cite{Jone89} gives 
\BEQ
\vm{R}_{i} (\theta) = \frac{\sinh \theta}{\sinh \eta} e_i
+ \frac{\sinh (\eta-\theta)}{\sinh \eta} \;\; ; \;\; q = e^{\eta}
\EEQ
which for our model (\ref{eq:B4},\ref{5:gl:ei}) leads to 
\BEQ
\vm{R}_{i} (\theta) = {\bf 1}_{1} \otimes \ldots \otimes {\bf 1}_{i-1}
\otimes {\cal R}_{i,i+1} 
\otimes {\bf 1}_{i+2} \otimes \ldots
\EEQ
with the non-vanishing elements of the matrix ${\cal R}_{i,i+1}$
\BEQ
{\cal R}_{i,i+1} := \frac{1}{\sinh \eta} \left(
\begin{array}{cccc}
\sinh (\eta-\theta) &  &  & \Omega \sinh \theta \\
 & e^{-\theta} \sinh \eta & e^{\eta} \sinh \theta & \\
 & e^{-\eta} \sinh \theta & e^{\theta} \sinh \eta & \\
 & & & \sinh (\eta+\theta) \end{array} \right)
\EEQ
The relations eq.~(\ref{3:gl:Hecke}) imply that these matrices satisfy
the spectral parameter-dependent braid-group relations
\BEA
\vm{R}_{i} (\theta) \vm{R}_{i\pm 1} (\theta+\theta') \vm{R}_{i} (\theta')
&=& \vm{R}_{i\pm 1} (\theta') \vm{R}_{i} (\theta+\theta')
\vm{R}_{i\pm 1} (\theta) \nonumber \\
\left[ \vm{R}_{i} (\theta) , \vm{R}_{j} (\theta') \right] &=& 0
\;\; ; \;\; |i-j| \geq 2
\EEA
which are equivalent to the Yang-Baxter equations.

%%----------------------------------------------------------------------------%%
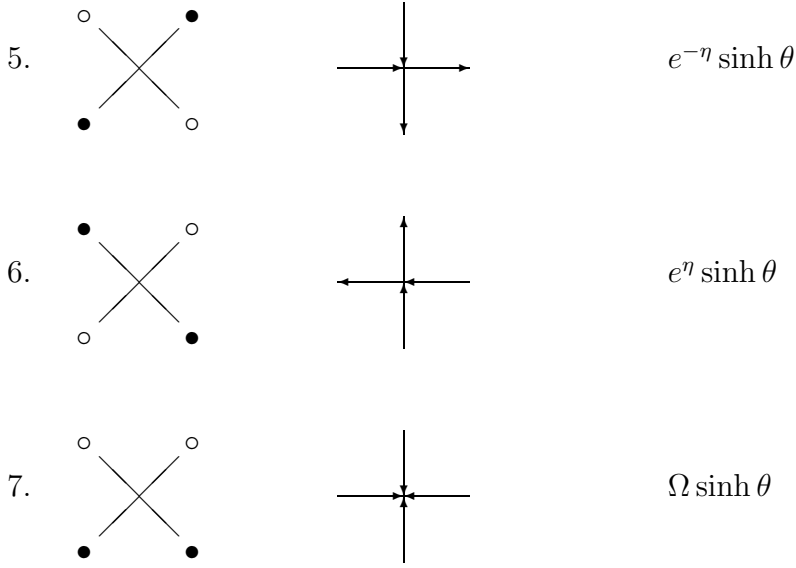
\begin{figure}[t]
\caption{Diffusion and pair-annihilation of particles in the seven-vertex 
model and their Boltzmann weights, for the vertices 5 to 7.\label{Abb2}}
\begin{picture}(300,80)(0,0)
 \put(250,50){$e^{-\eta} \sinh \theta$}
%
% Das Zentrum ist bei (150,50)
%
 \put(150,50){\vector(1,0){25}}
 \put(125,50){\vector(1,0){25}}
 \put(150,75){\vector(0,-1){25}}
 \put(150,50){\vector(0,-1){25}}

 \put(50,50){\line(1,1){15}}
 \put(50,50){\line(-1,1){15}}
 \put(50,50){\line(-1,-1){15}}
 \put(50,50){\line(1,-1){15}}

 \put(67,67){$\bullet$}
 \put(26,67){$\circ$}
 \put(26,26){$\bullet$}
 \put(67,26){$\circ$}

 \put(0,50){$5.$}

\end{picture}

\begin{picture}(300,80)(0,0)
 \put(250,50){$e^{\eta} \sinh \theta$}
%
% Das Zentrum ist bei (150,50)
%
 \put(175,50){\vector(-1,0){25}}
 \put(150,50){\vector(-1,0){25}}
 \put(150,50){\vector(0,1){25}}
 \put(150,25){\vector(0,1){25}}

 \put(50,50){\line(1,1){15}}
 \put(50,50){\line(-1,1){15}}
 \put(50,50){\line(-1,-1){15}}
 \put(50,50){\line(1,-1){15}}

 \put(67,67){$\circ$}
 \put(26,67){$\bullet$}
 \put(26,26){$\circ$}
 \put(67,26){$\bullet$}

 \put(0,50){$6.$}

\end{picture}

\begin{picture}(300,80)(0,0)
 \put(250,50){$\Omega \sinh \theta$}
%
% Das Zentrum ist bei (150,50)
%
 \put(175,50){\vector(-1,0){25}}
 \put(125,50){\vector(1,0){25}}
 \put(150,75){\vector(0,-1){25}}
 \put(150,25){\vector(0,1){25}}

 \put(50,50){\line(1,1){15}}
 \put(50,50){\line(-1,1){15}}
 \put(50,50){\line(-1,-1){15}}
 \put(50,50){\line(1,-1){15}}

 \put(67,67){$\circ$}
 \put(26,67){$\circ$}
 \put(26,26){$\bullet$}
 \put(67,26){$\bullet$}
 
 \put(0,50){$7.$}

\end{picture}
\end{figure}
%%----------------------------------------------------------------------------%%

In a $2D$ vertex model with vertex configurations labelled
by $(k,\ell,m,n)$, the Boltzmann weights $S_{\ell,m}^{kn}$ are obtained from
\BEQ
\vm{R}_{i} (\theta) = S_{\ell,m}^{k,n} {\bf 1}_{1} \otimes
\ldots \otimes {\bf 1}_{i-1} \otimes E^{mk} \otimes E^{n\ell}
\otimes {\bf 1}_{i+2} \otimes \ldots
\EEQ
This implies that the vertex model associated to eq.~(\ref{eq:B4})
is a {\em seven-vertex model}. In a vertex model, arrows are attached to 
the bonds of a square lattice \cite{Baxt82}. 
In the stochastic model, we associate a
particle ($\bullet$) with an arrow pointing up/right and no particle ($\circ$)
with an arrow pointing down/left. In figure~\ref{Abb2} we list
together the chemical reactions, the vertex configurations and their Boltzmann
weights. The vertices usually labelled 1 to 4 correspond to no reaction and are
not shown (see \cite{Alca94}). Vertices 5 and 6 correspond to diffusion 
to the right and to the left and vertex 7 to pair-annihilation. In the leftmost 
column of figure~\ref{Abb2}, the state of the particles before the reaction is 
given as the lower pair of symbols while the state after the reaction is given 
by the upper pair of symbols. The middle column gives the corresponding vertex
configuration and the right column the Boltzmann weight. 
The Hamiltonian eq.~(\ref{eq:B4}) may be 
recovered from $H = -\left.\frac{\D}{\D\theta}\ln T(\theta)\right|_{\theta=0}$.

%%%%%%%%%%%%%%%%%%%%%%%%%%%%%%%%%%%%%%%%%%%%%%%%%%%%%%%%%%%%%%%%%%%%%%%%%%%%%%%%
\section{Further applications}
%%%%%%%%%%%%%%%%%%%%%%%%%%%%%%%%%%%%%%%%%%%%%%%%%%%%%%%%%%%%%%%%%%%%%%%%%%%%%%%%

We have studied in some detail the pair-annihilation process and its 
integrability. Still, extracting explicitly
information about the long-time behaviour (or the steady-state in more
complicated models) is not yet trivial. In this section, 
we briefly review some approaches which may be useful. \\

\noindent {\bf A) Spectral integrability}. 
We have already seen that in certain cases,
the quantum Hamiltonian $H= H_0 + H_1$ such that $\mbox{\rm spec}(H) =
\mbox{\rm spec}(H_0)$, independently of the precise form of $H_1$. It may 
happen that although $H_0$ is integrable, $H$ is not. Such a model is said
to be {\em spectrally integrable}. If only binary interactions are present, 
it is convenient to express $H$ in terms of a two-site matrix $H_{i,i+1}$ 
acting on the sites $i$ and $i+1$ 
\BEQ \label{6:gl:Hzwei}
H = \sum_{j=1}^{L} {\bf 1}_1 \otimes \cdots \otimes {\bf 1}_{j-1} \otimes
H_{j,j+1} \otimes {\bf 1}_{j+2} \cdots \otimes {\bf 1}_L
\EEQ
For the model (\ref{4:gl:HDacd}) with left-right symmetry, that is $q=1$, 
we have 
\BEQ \label{6:gl:Hmat}
H_{j,j+1} = D \left( \begin{array}{cccc}
0 &  -\delta &  -\delta & -2\alpha \\
0 &  1+\delta & -1 & -\gamma \\
0 & -1 &  1+\delta & -\gamma \\
0 &  0 &  0 & 2(\alpha+\gamma) \end{array} \right)
\EEQ
One can always rescale time such that $D=1$. Then the parameters of the XXZ 
chain $H_0=H_{\rm XXZ}$ 
become $\Delta=1+\delta-\gamma-\alpha$ and $h=\alpha+\gamma$. 

In eqs.~(\ref{6:gl:Hzwei},\ref{6:gl:Hmat}) $H$ is only integrable for 
$\delta=0$. On the other hand, the special case $\alpha=0,\gamma=\delta$ 
simply corresponds to the radioactive decay of diffusively moving particles. 
While for $\delta=0$, one is back to the critical Pokrovsky-Talapov line
$h+\Delta=1$, the associated quantum spin chain has a frozen ground state for
$\delta=0$. The one-particle and two-particle correlators 
$C_1(t)=\sum_{j=1}^{L} C_1(j;t)$ and $C_{2;n}(t) =\sum_{j=1}^{L} C(j,j+n;t)$
with $n$ fixed, see eq.~(\ref{OneTwo}), only imply the sectors with $r=1$ and 
$r=2$ particles of $H_{\rm XXZ}$, respectively. Their long-time behaviour is
easily worked out from the results of section~4 and is collected in
table~\ref{tab2} \cite{Henk97}. 
Of course, we implicitly assume that the corresponding
amplitudes do not accidentally vanish. 
At first sight, one might have expected a simple
exponential factor $e^{-2 k\delta t}$ for the $k$-point correlator $C_k$ and
we already observe that eventual algebraic prefactors are not
readily predicted from the spectrum of $H$ alone. The more complicated
form of the relaxation time for $\delta>\alpha+\gamma$ comes from a bound
state in the two-particle sector of $H_{\rm XXZ}$, with energy
$4\delta+4-2\Delta-2/\Delta$, see \cite{Baxt82,Gaud83}. More general
initial conditions are discussed in \cite{Henk97}.\\ 

\begin{table} 
\begin{center}
\begin{tabular}{|c|cc|} \hline 
$\delta$ & $C_1(t)$ & $C_{2;n}(t)$  \\ \hline
0 & $t^{-1/2}$ & $t^{-3/2}$  \\
$<\alpha+\gamma$ & $\exp(-2\delta t)$ & $t^{-1/2} \exp(-4\delta t)$  \\
$>\alpha+\gamma$ & $\exp(-2\delta t)$ & $\exp{\left[ -4\delta t +
2(\Delta+1/\Delta -2)t\right]}$ \\ 
\hline
\end{tabular}
\caption[Generics]{Generic long-time behaviour of the one-point and two-point
correlators $C_1(t)$ and $C_{2;n}(t)$ for $n$ finite, 
in the system eq.~(\ref{6:gl:Hmat}) and with a translation-invariant 
initial state and a finite initial particle-density. 
\label{tab2}}
\end{center} 
\end{table}

\noindent {\bf B) Similarity transformations}. 
In trying to extract explicit information
on certain reaction-diffusion systems, the integrability of the quantum
Hamiltonian $H$ plays a central r\^ole. Since it is difficult to
realize the constraints (\ref{2:gl:StoBed}) of stochasticity and integrability 
at the same time, it is of interest to see whether there exist systematic 
transformations of an integrable quantum Hamiltonian $H$ towards a new 
stochastic Hamiltonian $\wit{H}$. 

Specifically, we shall consider the transformation 
\BEQ \label{6:gl:BTrans}
\wit{H} = {\cal B} H {\cal B}^{-1} \;\; , 
\;\; {\cal B} = \bigotimes_{j=1}^{L} B_j
\EEQ
where $B_j={\bf 1}\otimes\ldots\otimes {\bf 1}
\otimes B \otimes {\bf 1}\otimes\ldots\otimes{\bf 1}$ is 
the transformation matrix $B$ acting on 
the site $j$. Then the systems described by $H$ and $\wit{H}$ are said to
be {\em similar} to each other. An interesting alternative, 
which has not yet been systematically studied, is to consider an 
{\em enantiodromy} transformation \cite{Schu00}
\BEQ
\wit{H} = {\cal B} H^{\rm T} {\cal B}^{-1}
\EEQ
where $H^{\rm T}$ is the transpose of $H$. 

From now on, we shall focus on translationally invariant
systems and consider periodic boundary conditions. 
The effect of the transformation $\cal B$ on $H$ is completely given by its
effect on the two-particle Hamiltonian $H_{j,j+1}$ 
in (\ref{6:gl:Hmat}). A {\em stochastic similarity transformation} arises if
both $H$ and $\wit{H}$ represent stochastic systems. For a simple example,
consider the symmetric annihilation-coagulation process 
eqs.~(\ref{6:gl:Hzwei},\ref{6:gl:Hmat}) with $\delta=0$. If
$C(t|\alpha,\gamma)=C_1(t)$ is the spatially averaged particle density
with the rates $\alpha$ and $\gamma$, respectively, a stochastic
similarity transformation shows that \cite{Kreb95,Henk95,Simo95} 
\BEQ \label{6:gl:Ctac}
C_1(t|\alpha,\gamma) = \frac{\alpha+\gamma}{2\alpha+\gamma}\, 
C_1(t|0,\alpha+\gamma)
\EEQ
Similar results hold for any $k$-point correlator $C_k(t)$. 
So far, explicit methods to find the time-dependent correlators are only
available for either the pure coagulation model $\alpha=0$ through
empty-interval methods (see below) or the pure annihilation model $\gamma=0$
through free-fermion techniques, see \cite{Lush86,Henk97,Schu00} and references 
therein. Eq.~(\ref{6:gl:Ctac}) allows to reduce any symmetric 
annihilation-coagulation process to pure coagulation, for any initial density 
$C_1(0|0,\gamma)$. This also explains the experimental results in 
table~\ref{tab:ReakDiffExp}. The known stochastic 
similarity transformations of the form (\ref{6:gl:BTrans}) 
leave the parameters $\Delta$ and $h$ of the XXZ 
chain invariant, but the results of
Proposition 2 suggest that a stochastic similarity transformation between 
systems with different values of $\Delta$ might exist. See \cite{Henk95} for
the extensions to $\delta>0$ and $q\ne 1$. 

Eq.~(\ref{6:gl:Ctac}) allows to recover the long-time behaviour 
$C(t|\alpha,\gamma)\sim t^{-1/2}$ from a simple heuristic argument. 
For pure coagulation, one particle always
remains, thus $C(\infty|0,\gamma)=1/L$ in the steady state. 
Therefore the steady-state density
$C(\infty|\alpha,\gamma)\sim L^{-1}$. On the other hand, from the spectrum
of $H$, the leading relaxation time $\tau=\Gamma^{-1}\sim L^2 \sim \xi^2$, where
$\xi$ is identified as the characteristic spatial length scale. Therefore
$C(\infty|\alpha,\gamma)\sim \xi^{-1}\sim \tau^{-1/2}$. The asserted 
time-dependent behaviour therefore might have been anticipated on 
dimensional grounds.\\ 

\noindent {\bf C) Free fermions}. 
For the pure annihilation model with $2\alpha=D_R+D_L$,
one has $\Delta=0$. In this case, $H$ may be diagonalized through a 
Jordan-Wigner transformation followed by a canonical 
transformation \cite{Lush86}. In order for this to work, $H$ may only contain
{\em pairs} of particle creation and annihilation operators. 
For space-independent reaction rates, the complete list of reaction-diffusion
process whose quantum Hamiltonian $\wit{H}$ is similar through 
(\ref{6:gl:BTrans}) to a free-fermion Hamiltonian $H$ 
is as follows \cite{Schu00,Henk97,Henk95} and shown in table~\ref{tab4}. 
Since the transformation (\ref{6:gl:BTrans}) is spatially local, these 
correspondences actually hold in any dimension, but free-fermion methods 
are only available in $1D$.  

\begin{table}
\begin{tabular}{|c|ccc|cc|}  \hline
model & \multicolumn{3}{c|}{reactions} & \multicolumn{2}{c|}{conditions}
\\ \hline
A & $\bullet\bullet\leftrightarrow\circ\circ$ &  &
   $\bullet\circ\leftrightarrow\circ\bullet$ & 
   \multicolumn{2}{c|}{$2(\alpha+\sigma)=D_L+D_R$} \\
B & $\bullet\bullet\rightarrow\circ\circ$ & 
   $\bullet\bullet\rightarrow\circ\bullet,\bullet\circ$ &
   $\bullet\circ\leftrightarrow\circ\bullet$ &
   \multicolumn{2}{c|}{$2\alpha+\gamma_L+\gamma_R=D_L+D_R$} \\
C & $\circ\circ\rightarrow\bullet\bullet$ & 
   $\circ\circ\rightarrow\circ\bullet, \bullet\circ$ &
   $\bullet\circ\leftrightarrow\circ\bullet$ &
   \multicolumn{2}{c|}{$2\sigma+\nu_L+\nu_R=D_L+D_R$} \\
D & $\bullet\bullet\rightarrow\bullet\circ, \circ\bullet$ &
   $\bullet\circ, \circ\bullet\rightarrow\bullet\bullet$ &
   $\bullet\circ\leftrightarrow\circ\bullet$ &
   $\gamma_L=D_L$ & $\gamma_R=D_R$ \\
E & $\circ\circ\rightarrow\bullet\circ, \circ\bullet$ &
   $\bullet\circ, \circ\bullet\rightarrow\circ\circ$ & 
   $\bullet\circ\leftrightarrow\circ\bullet$ &
   $\nu_L=D_L$ & $\nu_R=D_R$ \\
F & $\bullet\circ, \circ\bullet\rightarrow\circ\circ$ &
   $\bullet\circ, \circ\bullet\rightarrow\bullet\bullet$ & &
   \multicolumn{2}{c|}{$\delta_R/\delta_L=\beta_R/\beta_L$} \\
G & \multicolumn{2}{c}{
   $\bullet\circ, \circ\bullet\leftrightarrow\bullet\bullet, \circ\circ$} & &
   \multicolumn{2}{c|}{$\left\{\begin{array}{cc} 
   \delta_{R}=\beta_{R} & \gamma_{R}=\nu_{L} \\ 
   \delta_{L}=\beta_{L} & \gamma_{L}=\nu_{R} \end{array}\right.$} \\
H & $\bullet\rightarrow\circ$ & $\circ\rightarrow\bullet$ & &
   \multicolumn{2}{c|}{$\left\{\begin{array}{c} 
   \delta_L=\delta_R=\gamma_L=\gamma_R \\ \beta_L=\beta_R=\nu_L=\nu_R
   \end{array} \right.$} \\ 
   \hline
\end{tabular}
\caption{Single-species processes with space-independent reaction rates and 
which are similar via (\ref{6:gl:BTrans}) to a free-fermion model. 
The reaction rates are defined in table~\ref{tab1}.\label{tab4}}
\end{table}

Of these models, only models A (diffusive pair-annihilation and creation, 
solved exactly in $1D$ in \cite{Lush86}),
G (kinetic Ising model with Glauber dynamics) and H 
(free decay and creation
of particles) are reversible and have an equilibrium steady-state. Their
quantum Hamiltonian is therefore similar to a symmetric matrix. The similarity
of the kinetic Ising model with Glauber dynamics (model G) to a free-fermion
model was obtained long ago through a duality transformation \cite{Sigg77} and
more recently as a true similarity transformation \cite{Henk95,Sant97}. This
suggests to study a more general type of relationship, based on domain-wall 
dualities, see \cite{Schu00,Sant97} for details. Models C and E are obtained 
by a particle-hole permutation $\bullet\leftrightarrow\circ$
from models B and D, respectively. Model B is the biased 
annihilation-coagulation process, while model D is the diffusive coagulation
process with arbitrary decoagulation. Finally, model F is the doubly
biased voter model (in space and in the preference between 
$\bullet$ and $\circ$) and in \cite{Agha00} some correlators are found from 
the free-fermion form. The physical behaviour of all these models can
be treated in a single calculation. For example, the mean particle-density
depends on a single parameter $h$ such that \cite{Henk97}
\BEQ
C_1(t) - C_1(\infty) \sim \left\{
\begin{array}{ll} t^{-1/2} & \mbox{\rm ~~;~~ if $h=1$} \\
t^{-3/2} \exp(-t/\tau) & \mbox{\rm ~~;~~ if $0<h<1$}
\end{array} \right.
\EEQ
where $C_1(\infty)$ is the steady-state density and 
$\tau=1/(4-4h)$ is the re\-la\-xa\-tion~time (see \cite{Schu00} and references
therein for more information on solved free-fermion models).  

This kind of analysis was generalized to find those reaction-diffusion systems
which are similar, via a transformation of the type (\ref{6:gl:BTrans}),  
to the XXZ chain \cite{Henk97}. While the full result
is too complex to be re-stated here, an interesting special case is given by 
the conditions
\BEA
\gamma_R+\beta_L+2\alpha+D_L &=& \nu_L+\delta_L+2\sigma+D_R 
\nonumber \\
\gamma_L+\beta_R+2\alpha+D_R &=& \nu_R+\delta_R+2\sigma+D_L
\EEA 
In this case the (usually infinite) hierarchy of 
equations of motion for the $k$-point particle-density correlators 
${\cal C}_k(t) =\,<\!n_1(t)\ldots n_k(t)\!>$ closes naturally, such that
$\dot{\cal C}_k(t)$ only depends on the ${\cal C}_{\ell}(t)$ with $\ell\leq k$. 
In principle the equations of motions for the ${\cal C}_k$ can then be solved 
iteratively \cite{Schu95}.\\

\begin{figure}[th]
\centerline{\epsfxsize=3.05in\epsfbox
{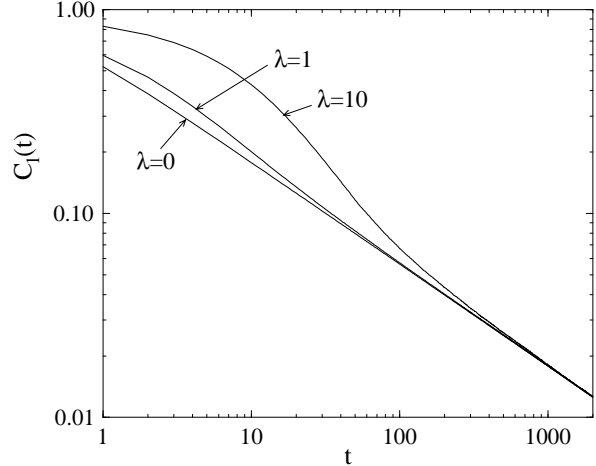}
}
\caption[Evolution of density]{Evolution of the mean particle density 
$C_1(t)$ in the symmetric coagulation model with the production reaction 
$\bullet\circ\bullet\rightarrow\bullet\bullet\bullet$ for several
production rates $\lambda$. For long times, the asymptotic behaviour is 
$C_1(t)\sim 1/\sqrt{t}$ for all values of $\lambda$ 
(after \protect{\cite{Henk01a}}).
\label{Bild3}}
\end{figure}

\noindent {\bf D) Partial integrability}. The previous sections have shown that 
constructing integrable stochastic systems which go beyond mere free diffusion
is a non-trivial exercice. One might wonder whether the condition of full
integrability is not too strong. After all, from a practical point of view
it would be enough to identify a set $\{Q_1,\ldots,Q_M\}$ of observables such 
that these satisfy a closed set of equations, say 
$\dot{Q}_i = f_i(Q_1,\ldots,Q_M)$, with $i=1,\ldots,M$. Such a
{\em partial integrability} may be enough for many practical needs. 
Indeed, such an approach is available through the {\em empty-interval method} 
\cite{Avra90,Avra00}. Consider a periodic chain with $L$ sites and 
lattice spacing $a$. Let $I_n(t)$ be the probability that at time $t$, 
$n$ consecutive sites are empty. Then the mean particle-density 
is \cite{Avra90,Kreb95}
\BEQ
C_1(t) =  \left( 1 - I_1(t) \right)/a
\EEQ
In order to illustrate the method, we consider the left-right symmetric
pure coagulation model and also take the free-fermion condition $\gamma=D$ of
model D in table~\ref{tab4}, but we now add a three-site production reaction
$\bullet\circ\bullet\rightarrow\bullet\bullet\bullet$ with rate $2D\lambda$
\cite{Henk01a}. The equations of motion for the $I_n(t)$ read
\BEA
\dot{I}_1(t) &=& 2D\left( I_0(t) - 2I_1(t) + I_2(t)\right) 
-2D\lambda \left( I_1(t) -2I_2(t)+I_3(t) \right) 
\nonumber \\
\dot{I}_n(t) &=& 2D\left( I_{n-1}(t) -2I_n(t)+I_{n+1}(t)\right) \;\; ;
\;\; 2\leq n\leq L-1
\label{6:gl:In}
\EEA
together with the boundary conditions $I_0(t)=1$ and $I_L(t)=0$ (assuming that
there is at least one particle in the system). The solution of these
equations is straightforward. For example, one obtains for the leading
relaxation time $\tau^{-1}=\Gamma=2D\pi^2 L^{-2}+O(L^{-4})$, in agreement
with the results of section~4. The effect of the production term is only
transient, as illustrated in figure~\ref{Bild3} for the mean density $C_1(t)$. 
For $\lambda=0$, $C_1(t)\sim 1/\sqrt{t}$ is of course expected from 
eqs.~(\ref{1:gl:exakt},\ref{6:gl:Ctac}). 
While the free-fermion condition $\gamma=D$ 
is essential for the method to work, we also see that the presence of the
production term poses no problem at all for the closure of the equations of 
motion (\ref{6:gl:In}).\footnote{This term cannot, e.g. by a similarity 
transformation, be turned into a term treatable by either free-fermion or 
full integrability methods.} 

%%----------------------------------------------------------------------------%%
%%----------------------------------------------------------------------------%%

Accepting the free-fermion condition $\gamma_{L,R}=D_{L,R}$, 
one can extend the treatment to the more general model D of table~\ref{tab4} 
and may even extend this further to include the processes
$\circ\circ\rightarrow\bullet\bullet$ and 
$\circ\circ\rightarrow\circ\bullet, \bullet\circ$ with rates 
$2\sig,\nu_R,\nu_L$, respectively \cite{Avra00,Avra90}. 
Let us call this system {\em model D'\/} which depends on the seven parameters
$D_{L,R},\beta_{L,R},\nu_{L,R},\sig$. 
In a remarkable paper \cite{Pesc94}, the idea of the empty-interval method was 
translated into the Hamiltonian formalism and several new sets of observables
were defined which generalize the variables $I_n(t)$ and lead again to closed 
equations of motion. It turns out that the spectrum of
relaxation times of model D' is given by the Hamiltonian of the 
Wannier-Stark ladder \cite{Pesc94}
\BEQ
H = - \sum_{n=-L}^{L} \left[ \sig_n^x\sig_{n+1}^x + \sig_n^y\sig_{n+1}^y 
+ (h+h' n)\sig_n^z\right]
\EEQ
where $h$ and $h'$ are constants. In this case, the couplings in $H$ are
space-dependent. The extension of the similarity/enantiodromy approach to this
more general setting remains to be done. Extensions of the empty-interval
method to interactions on more than two sites are studied in \cite{Khor03}. 

While the empty-interval method as such does not work for the pair-annihilation
process, the method has been generalized recently \cite{Mass01a}. 
We briefly explain the idea using the left-right symmetric pair-annihilation 
process with the free-fermion condition $\alpha=D$ (model A or B) as example. 
Let $G_n(t)$ be the probability that at time $t$, one has on $n$ consecutive
sites an even number of particles. The mean particle density is 
$C_1(t) = (1-G_1(t))/a$. Furthermore, let $F_n(t)$ ($H_n(t)$) be the 
probability that a segment of $n$ consecutive sites with an even (odd) 
number of particles is followed by the presence of a particle at the 
$(n+1)^{\rm th}$ site. From the relations
\BEQ
2 F_n(t) = (1 - G_1)+(G_n-G_{n+1}) \;\; , \;\;
2 H_n(t) = (1 - G_1)-(G_n-G_{n+1}) 
\EEQ
and the boundary condition $G_0(t)=1$, the equations of motion
\BEQ
\dot{G}_n(t) = 2D \left( F_{n-1}-H_{n-1} +H_n-F_n\right) = 2D \left(
G_{n-1}(t) -2G_n(t)+G_{n+1}(t)\right)
\EEQ
follow. They can be solved
by standard methods. Reaction terms para\-me\-tri\-zed by 
$\sig,\nu_{L,R},\beta_{L,R}$ (see table~\ref{tab1}) and even the
reaction $\circ\bullet\circ\rar\bullet\bullet\bullet$ 
can be added \cite{Mass01a}. Correlators are studied in \cite{Mass01b}.

In view of the practical success of these techniques it is perhaps not 
completely futile to ask whether there might a be a systematic way to 
identify these `empty-interval' or related variables~? \\ 

\noindent {\bf E) Multi-species models}. We consider
chains with $N$ states per site. One of them is taken to be the empty
site ($\circ$) and the other states are labelled $A_n$, $n=1,\ldots,N-1$. 
Finding integrable stochastic systems becomes more difficult when $N$ 
increases. Several examples were found \cite{Alca93} 
through the quotients $(P,M)H_{L-1}(q)$ as realized through
the Perk-Shultz model. They are collected in table~\ref{tab5}. The following
conventions apply. 
\begin{enumerate}
\item For the first reaction in all models and the second reaction
in models $\mathfrak{B,C,D}$ (with $n<m$ understood) the reaction to the 
right (left) occurs with rate $\Gamma_R$ ($\Gamma_L$). 
\item For models 
$\mathfrak{E,F}$ the sum $r+s$ has to be taken $\bmod N$. 
If in this case $r+s=0\bmod N$, the rate is $\Gamma_L+\Gamma_R$. If
$r+s\ne 0\bmod N$ as well as for the third reaction in models
$\mathfrak{C,D}$ the rate is $\Gamma_R$. In model
$\mathfrak{D}$, it is also assumed that in the third reaction, 
pairs $(r,s=r\pm 1)$ never have an element in common. 
If the products on the right are
interchanged, (e.g. $A_1A_1\rightarrow A_1\circ$ in model $\mathfrak{C}$), 
the rate is $\Gamma_L$. 
\item In the third reaction in model $\mathfrak{F}$ the
rates are $\Gamma_{\pm}$, respectively such that 
$\Gamma_{+}+\Gamma_{-}=\Gamma_L+\Gamma_R$.
\end{enumerate} 

One defines 
$D=\sqrt{\Gamma_L\Gamma_R}=1$ and $q=\sqrt{\Gamma_R/\Gamma_L}$. 
The Hecke algebra quotient $(P,M)H_{L-1}(q)$ according to the realization as 
a Perk-Shultz quantum chain 
eq.~(\ref{3:gl:PS}) \cite{Shul81,Alca93} is also indicated. 

\begin{table}
\begin{tabular}{|c|ccc|c|} \hline
model & \multicolumn{3}{c|}{reactions} & quotient \\ \hline
$\mathfrak{A}$ & $A_n\circ\leftrightarrow\circ A_n$ & & & ($2,0$) \\
$\mathfrak{B}$ & $A_n\circ\leftrightarrow\circ A_n$ & 
$A_n A_m\leftrightarrow A_m A_n$ & & ($N,0$) \\
$\mathfrak{C}$ & $A_n\circ\leftrightarrow\circ A_n$ & 
$A_n A_m\leftrightarrow A_m A_n$ &
$A_1 A_1\rightarrow\circ A_1$ & ($N-1,1$) \\
$\mathfrak{D}$ & $A_n \circ\leftrightarrow\circ A_n$ & 
$A_n A_m\leftrightarrow A_m A_n$ &
$A_r A_r\rightarrow A_{r\pm 1} A_r$ & ($N-2,2$) \\
$\mathfrak{E}$ & $A_n \circ\leftrightarrow\circ A_n$ &
$A_r A_s\rightarrow\circ A_{r+s}$ & & ($1,1$) \\
$\mathfrak{F}$ & $A_n \circ\leftrightarrow\circ A_n$ &
$A_r A_s\rightarrow\circ A_{r+s}$ &
$A_n A_n\rightarrow A_{n\pm 1} A_{n\pm 1}$ & ($2,1$) \\ \hline
\end{tabular}
\caption[Integrable processes]{Some integrable reaction-diffusion processes of 
$N-1$ species and their Hecke algebra quotient \protect{\cite{Alca93}}, 
see text for the definition of the rates. 
\label{tab5}}
\end{table}

From table~\ref{tab5} and Theorem 3 we see that the simple diffusion model
$\mathfrak{A}$ has, up to degeneracies, the same spectrum as the XXZ chain
used in section~4 to described biased diffusion of a single species of 
particles $\bullet$. In the same way, the spectrum of model $\mathfrak{E}$ is,
up to degeneracies, the same as the one found for pair-annihilation in
section~4, with $2\alpha=D_L+D_R$. 

For illustration, we briefly consider model $\mathfrak{E}$ for $N=3$. Each site
may contain either a particle of type $A$ or $B$ or be empty ($\circ$). Single
particles may diffuse to the right $A\circ\rightarrow\circ A$,  
$B\circ\rightarrow\circ B$ with a rate $\Gamma_R$ or similarly to the left with
rate $\Gamma_L$. On encounter, between like particles the reactions
$AA\rightarrow\circ B$ and $AA\rightarrow B\circ$ occur with rates $\Gamma_R$
and $\Gamma_L$ respectively and similarly $BB\rightarrow\circ A, A\circ$. Two
unlike particles react $AB\rightarrow\circ\circ$ with rate $\Gamma_L+\Gamma_R$. 
In the left-right symmetric case, the identity of the spectra of 
$H_{(\mathfrak{E})}$ and the one of pair-annihilation, up to degeneracies, 
was checked directly \cite{Alca94}. Furthermore, in the spirit of the
empty-interval method, a closed system of equations of motion was found, 
whose solution leads to the mean particle-densities 
$\bar{n}_A(t)\sim\bar{n}_B(t)\sim t^{-1/2}$ \cite{Priv92}. 

In \cite{Alim02}, 
Bethe-ansatz solutions of the master equation for $N$-species models with
particle-numbers conservation are
studied. In particular, model $\mathfrak{B}$ with $N=3,5$ was rediscovered. 
The models in \cite{Alim02} are found from solutions of quantum Yang-Baxter 
equations. Further study might reveal a relationship to
diffusion algebras \cite{Isae01}, see below.  

For periodic boundary conditions and $N>2$, the diffusion bias leads after a 
similarity transformation to a generalized Dzialoshinsky-Moriya
interaction \cite{Dahm95}.   
A sufficient criterion for integrability was derived in an attempt to look 
more systematically for integrable many-species models \cite{Popk02}.

Finally, a different generalization from section~4 is to consider integrable 
stochastic models on ladders \cite{Albe01}, rather than chains. \\

\noindent {\bf F) Diffusion algebras}. 
For certain integrable systems, there exist
algebraic methods which allow to find the steady-state $\ket{s}$ such as the
celebrated matrix product states \cite{Derr93b,Derr93a}. Time-dependent 
problems are treated in \cite{Popk03}. 

Behind this seemingly technical and {\it ad hoc} method there is a new 
and general mathematical structure. 
We shall explain here the main idea using reaction-diffusion systems 
with $N$ states per site labelled by $A_n$, $n=0,1,\ldots,N-1$ 
(where $A_0=\circ$) moving on a periodic chain with $L$ sites, but
generalizations to different boundary conditions are possible. The allowed
reactions are $A_nA_m\rightarrow A_mA_n$ with rate $g_{nm}$ (in particular,
model $\mathfrak{B}$ from table~\ref{tab5} is a special case of this). 
The un-normalized steady-state probability distribution is \cite{Derr93b}
\BEQ
P_s(\vec{\sig}) = P(\sig_1,\ldots,\sig_L) = \mbox{\rm Tr}\left(
{\cal D}_{\sig_1} {\cal D}_{\sig_2}\cdots {\cal D}_{\sig_L}\right)
\EEQ
where the matrices ${\cal D}_{\sig}$ satisfy the quadratic relations
\BEQ
g_{\sig\rho}{\cal D}_{\sig}{\cal D}_{\rho} - 
g_{\rho\sig}{\cal D}_{\rho}{\cal D}_{\sig} = 
x_{\rho}{\cal D}_{\sig} - x_{\sig}{\cal D}_{\rho}
\EEQ
where $\sig<\rho$ and $\sig,\rho\in\{1,\ldots,N\}$ and 
$g_{\sig\rho}\in\mathbb{R}\backslash\{0\}$, $g_{\rho\sig}\in\mathbb{R}$ and the
$x_{\sig}$ are complex parameters. If in addition the set $\cal A$ of 
generators ${\cal D}_{\sig}$ admits a linear PBW basis of ordered monomials
${\cal D}_{\sig_1}^{k_1}{\cal D}_{\sig_2}^{k_2}\cdots{\cal D}_{\sig_n}^{k_n}$
with $\sig_1>\sig_2>\ldots>\sig_n$ and $k_n\in\mathbb{N}$, $\cal A$ is called
a {\em diffusion algebra} \cite{Isae01}. 

These conditions imply certain constraints on the $g_{\sig\rho}$ and the 
$x_{\sig}$, quite analogously to the Jacobi identities of a Lie algebra. 
These constraints can be fully solved and a classification of
all diffusion algebras for $N$ species is obtained \cite{Isae01,Pyat03}. The
representation theory of $N$-species diffusion algebras
is just getting started, see \cite{Twar02}.

%%%%%%%%%%%%%%%%%%%%%%%%%%%%%%%%%%%%%%%%%%%%%%%%%%%%%%%%%%%%%%%%%%%%%%%%%%%%%%%%
\section{Outlook: local scale invariance}
%%%%%%%%%%%%%%%%%%%%%%%%%%%%%%%%%%%%%%%%%%%%%%%%%%%%%%%%%%%%%%%%%%%%%%%%%%%%%%%%

We finish with a discussion on how the scale invariance of many 
reaction-diffusion systems might be turned into a dynamical symmetry. 
For example,
the symmetric pair-annihilation process is on the Pokrovsky-Talapov 
critical line. One has the covariance 
\BEQ
<\!n(r_1,t_1)\ldots n(r_p,t_p)\!> \; =\, b^{-(x_1+\cdots+x_p)}
<\!n(r_1',t_1')\ldots n(r_p',t_p')\!>
\EEQ
of the $p$-point correlators 
under the dilatation
$r\to r'=b r$, $t\to t'=b^z t$ of the space and time coordinates $r,t$ 
respectively, where $z$ is the dynamical exponent and $x_{1},\ldots x_p$ 
are scaling dimensions. In the cases at hand, $z=2$.  

This is reminiscent of the situation at equilibrium critical points. In those
systems, it is known that under fairly general conditions, the covariance of
the $p$-point correlators under global scale transformations 
$\vec{r}\to b \vec{r}$ can be extended to {\em conformal} transformations. 
In addition, in two dimensions the energy-momentum tensor of local conformally
invariant field theories becomes an analytic function $T=T(\mathfrak{z})$ of 
the complex coordinate $\mathfrak{z}$ such that not only $T(\mathfrak{z})$ 
itself, but also all powers $T^n(\mathfrak{z})$, $n=1,2,3\ldots$, 
are conserved \cite{Bela84,Zamo89}. 
This signals the integrability of $2D$ conformally invariant field theories. 
{\it Is it possible to generalize the space-time dilatations encountered for 
critical reaction-diffusion systems in a similar way~?}

This question has been recently addressed in the context of kinetic 
spin-systems \cite{Henk01b,Henk02}. 
We have already seen above that the kinetic Ising model with
Glauber dynamics \cite{Glau63} may be obtained through a similarity 
transformation of the quantum Hamiltonian from a certain single-species
reaction-diffusion system, see model G from table~\ref{tab4}. 
We now concentrate on this system. 
In the Glauber-Ising model the transition
rates in the master equation are chosen such that the steady-state $\ket{s}$
is given by the equilibrium probability distribution 
$P_s(\vec{\sig})\sim e^{-{\cal H}[\vec{\sig}]/T}$ with the classical Ising 
model Hamiltonian ${\cal H}=-\sum_{(i,j)} \sig_i \sig_j$ where $T$ is the 
temperature. Glauber dynamics may be realized through the discrete-time 
heat-bath rule $\sig_i(t)\to\sig_i(t+1)$ such that
\BEQ \label{7:gl:heat}
\sig_i(t+1) = \pm 1 \mbox{\rm ~~ with probability $\frac{1}{2}
\left[ 1\pm \tanh(h_i(t)/T)\right]$}
\EEQ
with the local field $h_i(t)=h+\sum_{j(i)} \sig_j(t)$. With the choice
(\ref{7:gl:heat}), the master equation can be solved exactly in 
$1D$ \cite{Glau63}. The time-dependent spin-spin 
correlators and their approach towards equilibrium are thus determined. 

%%----------------------------------------------------------------------------%%
\begin{figure}[t]
\centerline{\epsfxsize=2.3in\ \epsfbox{
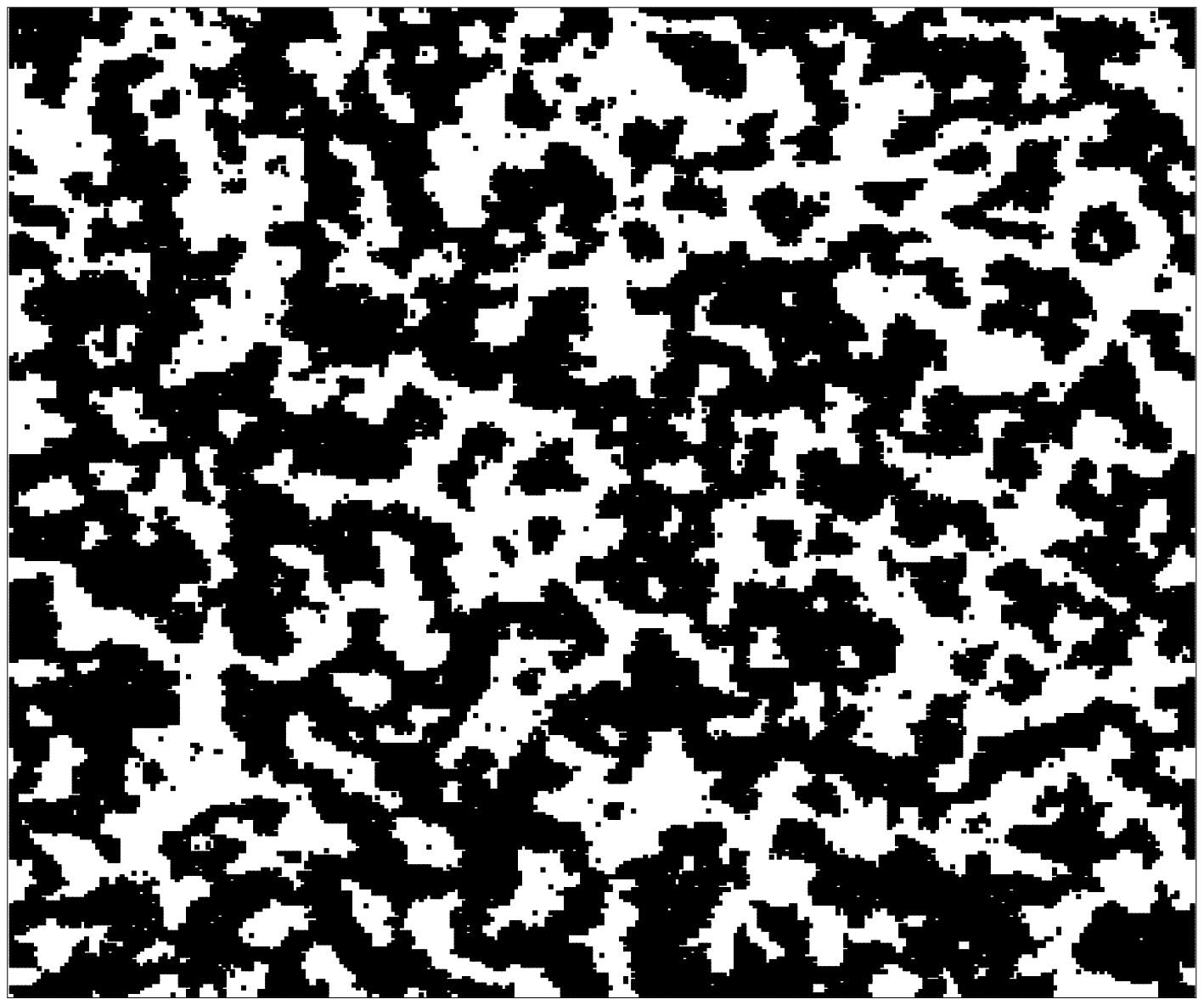} ~~
\epsfxsize=2.3in\epsfbox{
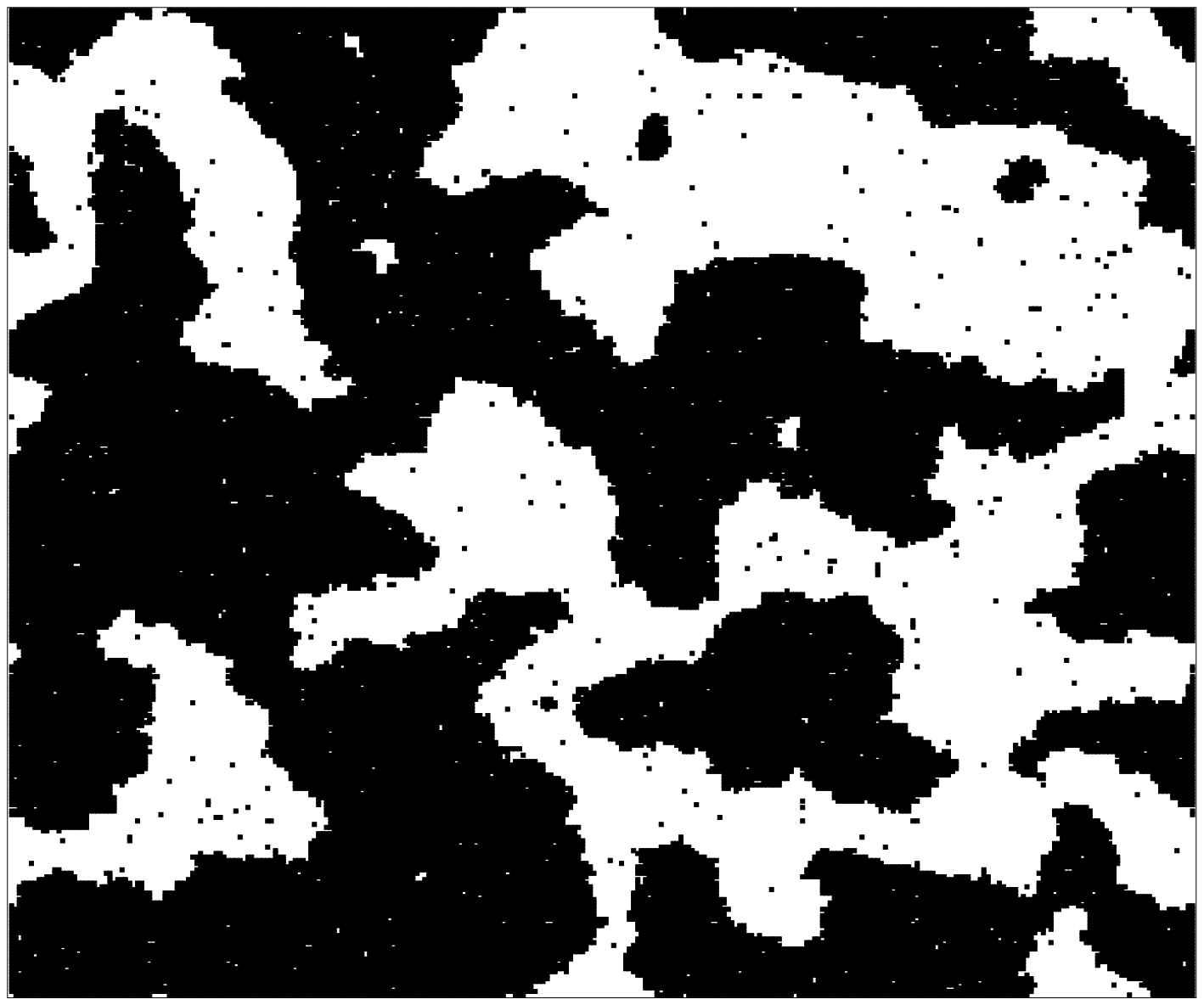}
}
\caption[Coarsening]{Snapshot of the coarsening of ordered domains in the 
$2D$ Glauber-Ising model, after a quench to $T=1.5<T_c$ from a totally 
disordered state and at times $t=25$ (left) and $t=275$ (right) after the 
quench.
\label{Bild4}}
\end{figure}
%%----------------------------------------------------------------------------%%

In contrast to equilibrium statistical mechanics, where fine-tuning the 
model parameters is needed to reach a critical point, dynamical scaling is 
often found to occur in large regions of the model's parameter space. 
For example, prepare the system initially in a disordered state and 
then quench the temperature to a final temperature $T<T_c$ below the critical
temperature $T_c>0$.\footnote{In the $1D$ Glauber-Ising model, $T_c=0$ leads to
certain modifications of the ageing as described from the point of view of
local scale invariance \cite{Pico03}.} 
Although the steady-state of the model is not critical, the
relaxation towards it occurs through domain coarsening and 
is very slow, the typical length scale varying with time as
$L(t)\sim t^{1/z}$, see figure~\ref{Bild4}. Typically the observables depend
algebraically on the time $t$ passed since the quench, 
see \cite{Cate00,Cugl02} for (a collection of) recent reviews. 
Here we concentrate on the two-time spatio-temporal response function 
$R(t,s;\vec{r})$ of the time-dependent spin $\sig_{\vec{r}}(t)$ at site
$\vec{r}$ with respect to an external magnetic field $h_{\vec{0}}(s)$ applied 
at the origin $\vec{0}$ at an earlier time $s<t$. 
Generically, two-time quantities such as $R(t,s;\vec{r})$ depend on
{\em both} times $t$ and $s$ and not merely on the difference $\tau=t-s$. This
breaking of time-translation invariance is called {\em ageing}. 

For ageing systems, an extension of dynamical scaling is possible and
allows to fix the form of the two-time response function. Specifically,
it can be shown that for a dynamical 
exponent $z=2$ \cite{Henk94a,Henk01b,Henk02}
\BEQ \label{7:gl:RR}
\!\!\!R(t,s;\vec{r}) = \left.\frac{\delta\langle\sig_{\vec{r}}(t)\rangle}
{\delta h_{{\vec{0}}}(s)}\right|_{h=0} \!= 
r_0 \left(\frac{t}{s}\right)^{1+a-\lambda_R/2} (t-s)^{-1-a} 
\exp\left[-\frac{\cal M}{2} 
\frac{\vec{r}^2}{t-s}\right]
\EEQ 
Here $a$ and $\lambda_R$ are  
non-equilibrium exponents to be determined which characterize the ageing
universality class \cite{Godr02}. Finally, $r_0$ and ${\cal M}$ are 
non-universal constants. We first present evidence that the response function 
of the Glauber-Ising model in $2D$ and $3D$ is indeed given by (\ref{7:gl:RR}). 
Then we discuss where this presumably exact result comes from. 

%%----------------------------------------------------------------------------%%
\begin{figure}[t]
\centerline{\epsfxsize=3.5in\ \epsfbox{
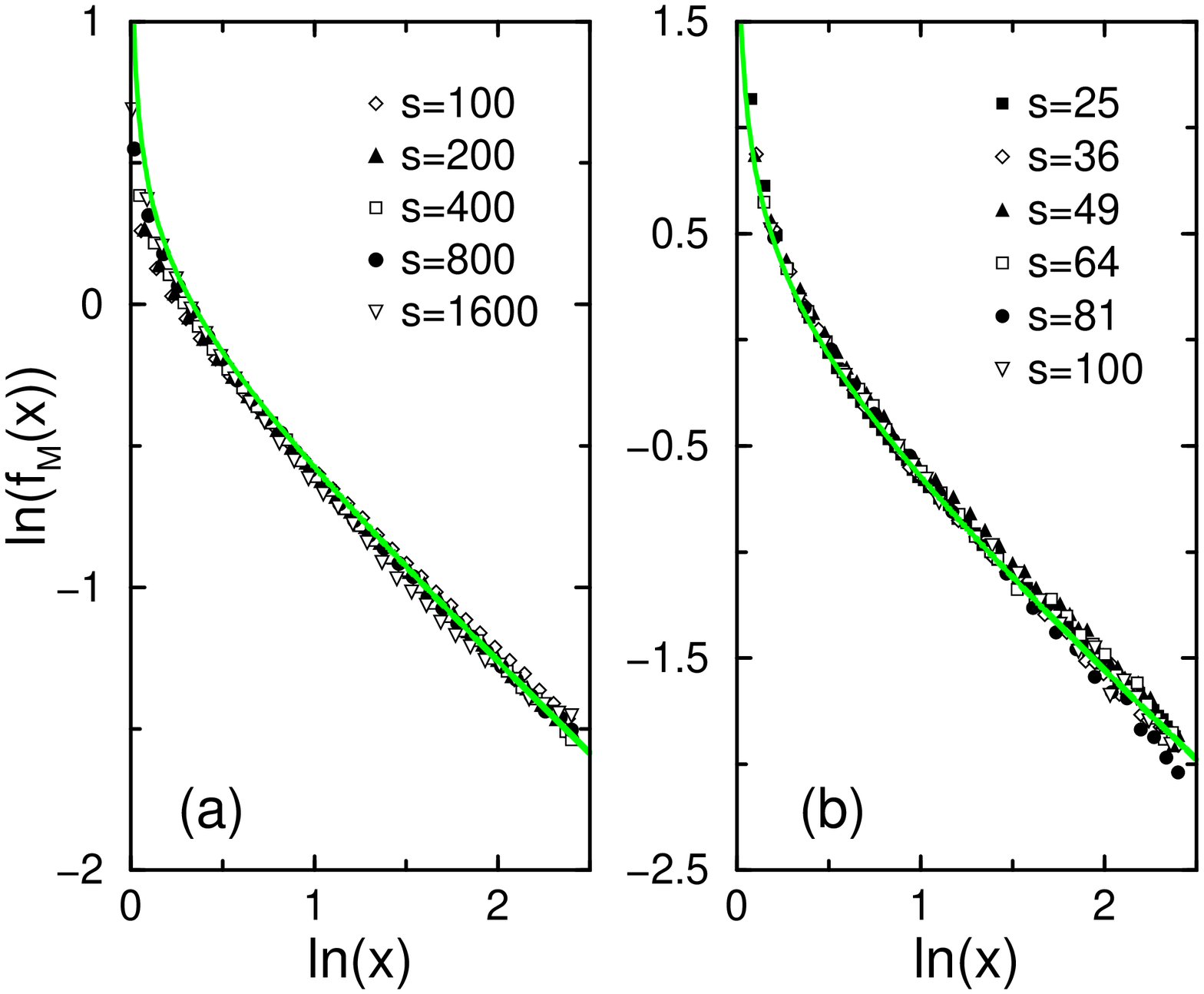} ~
\epsfxsize=1.88in\epsfbox{
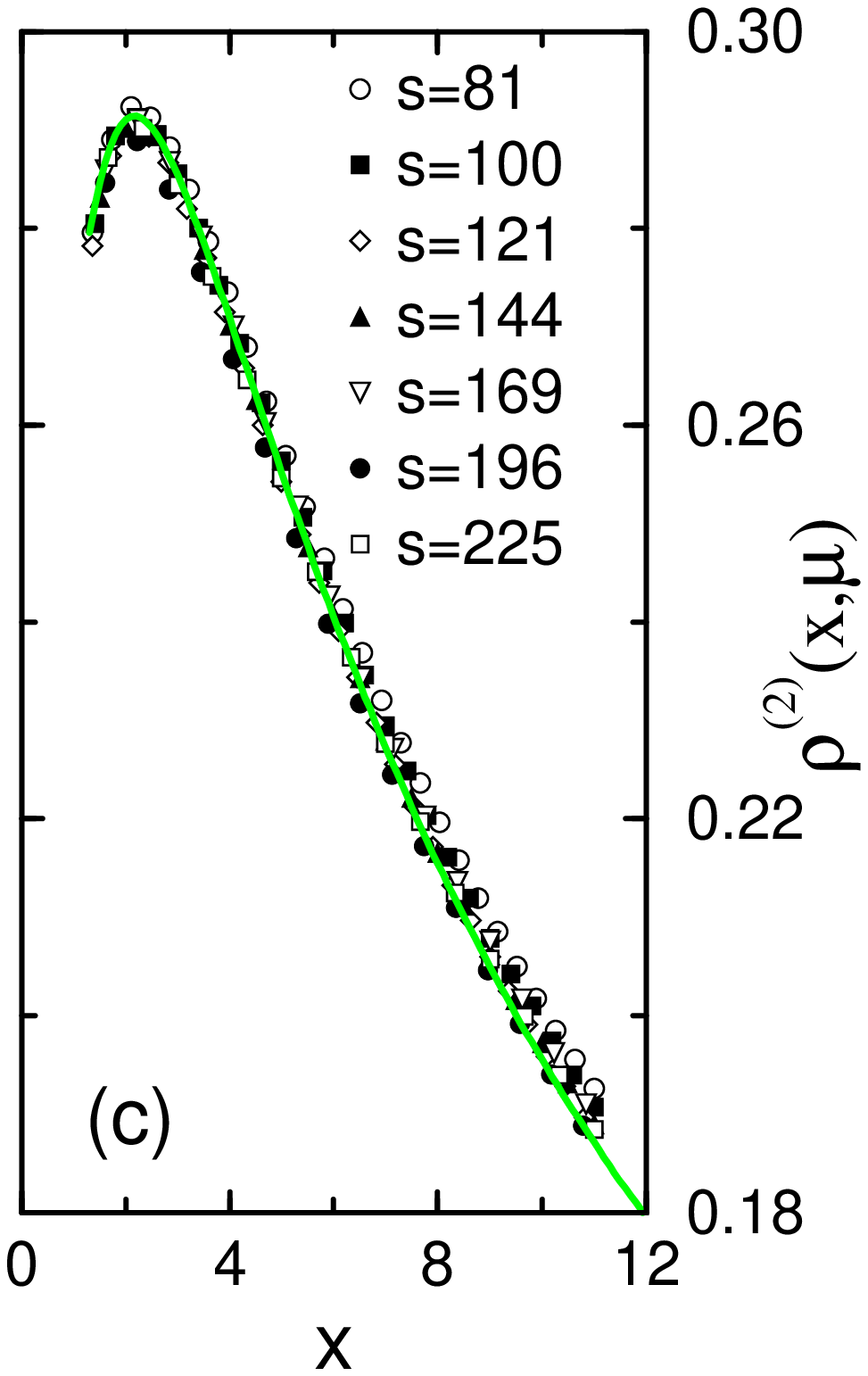}
}
\caption[Space-time response]{Scaling form of the integrated magnetic response
in the Glauber-Ising model as a function of $x=t/s$ below criticality. The
symbols correspond to different waiting times $s$. The integrated autoresponse 
is shown (a) $2D$ at $T=1.5$ and (b) $3D$ at $T=3$. An example of the 
integrated spatio-temporal response in $2D$ at $T=1.5$ and with $\mu=2$ is 
shown in (c). The full curves are obtained from (\ref{7:gl:RR}). 
After \protect{\cite{Henk02b}}.
\label{Bild5}}
\end{figure}
%%----------------------------------------------------------------------------%%

We first consider the autoresponse $R(t,s)=R(t,s;\vec{0})$. 
While $R$ itself is too noisy to be measured directly, integrated response
functions are accessible through simulations, see \cite{Henk02b} and
references therein for the details which we skip over here. 
For the example, the integrated
autoresponse 
\BEQ
\rho(t,s)=\int_{0}^{s} \!\D u\, R(t,u)\sim s^{-a} f_M(t/s)
\EEQ
is relatively easy to measure, whereas the scaling function $f_M(x)$ can be
calculated explicitly from (\ref{7:gl:RR}). In the Glauber-Ising model,
the exponent $a=1/z=1/2$ (see \cite{Henk02a} for a detailed discussion) 
and $\lambda_R\simeq 1.26$ and $1.6$
in $2D$ and $3D$, respectively. In figure~\ref{Bild5}ab, the scaling function
$f_M(x)$, as obtained from large-scale simulations, 
is shown for several values of
the waiting time $s$. In both two and three dimensions, a nice scaling behaviour
is found and the form of the scaling function agrees very well with the
prediction from eq.~(\ref{7:gl:RR}). Next, the $\vec{r}$-dependence of
$R(t,s;\vec{r})$ is 
tested by measuring the spatio-temporally integrated response
$\int_{0}^{s} \!\D u\int_{0}^{\sqrt{\mu s}} \!\D r\, r^{d-1} R(t,u;\vec{r}) 
\sim s^{d/2-a} \rho^{(2)}(t/s,\mu)$, where $\mu$ is a control 
parameter. We stress that the scaling function $\rho^{(2)}$ 
does not contain any more free non-universal parameter \cite{Henk02b}. As an
example, we compare in figure~\ref{Bild5}c data from $2D$ taken with $\mu=2$
with eq.~(\ref{7:gl:RR}). Besides the expected scaling, the functional form of
the scaling function neatly follows the prediction. We stress that the
position, the height and the width of the maximum of $\rho^{(2)}$ in 
figure~\ref{Bild5}c are completely fixed. Similar results have been
obtained for other values of $\mu$ and in $3D$ as well. This provides strong
evidence that eq.~(\ref{7:gl:RR}) is exact, at least in this model 
\cite{Henk02b}. Tests of (\ref{7:gl:RR}) in different universality classes are 
described in \cite{Henk02}. 
 
\begin{figure}[th]
\centerline{\epsfxsize=1.5in\ \epsfbox{
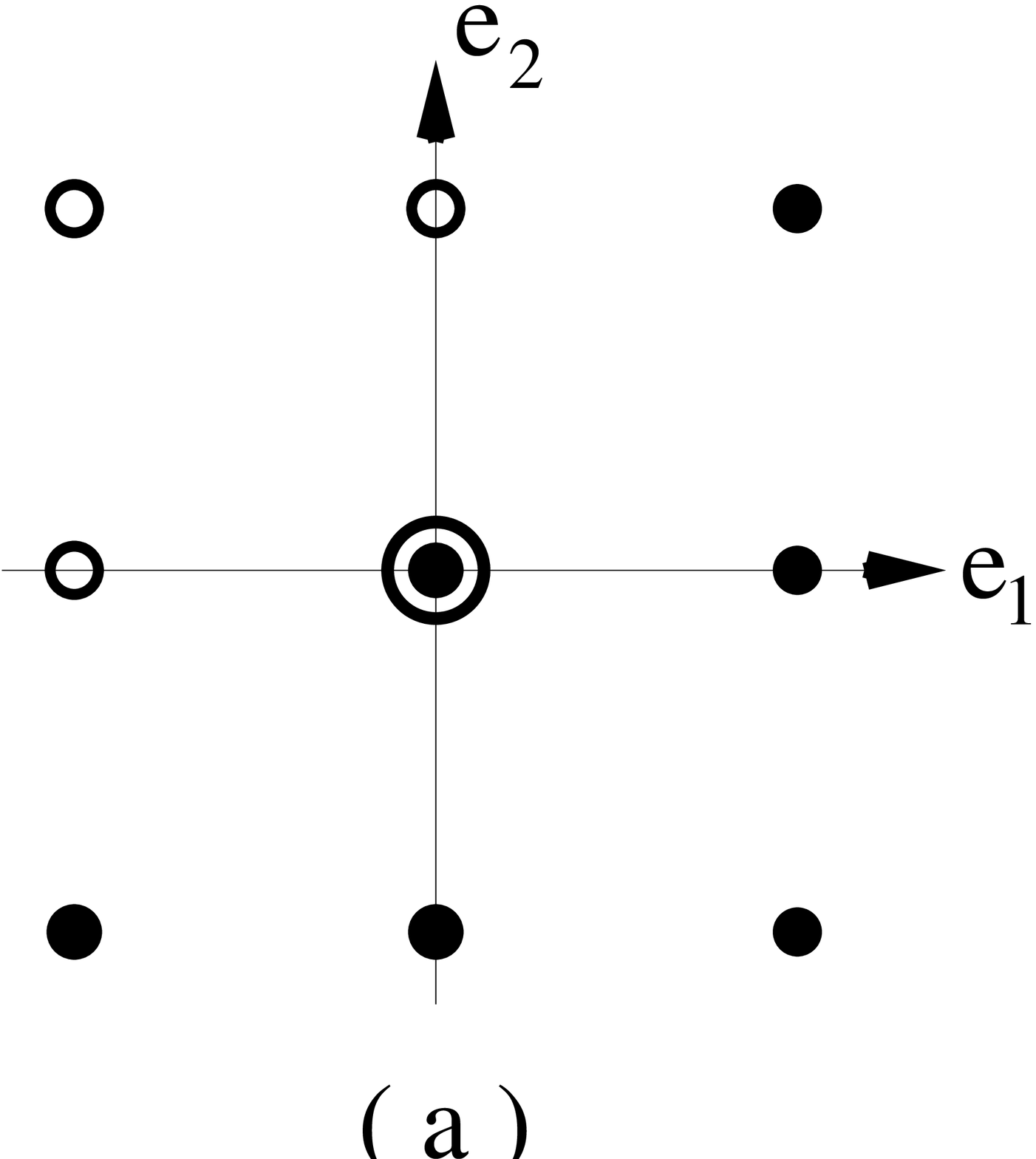} ~
\epsfxsize=1.5in\epsfbox{
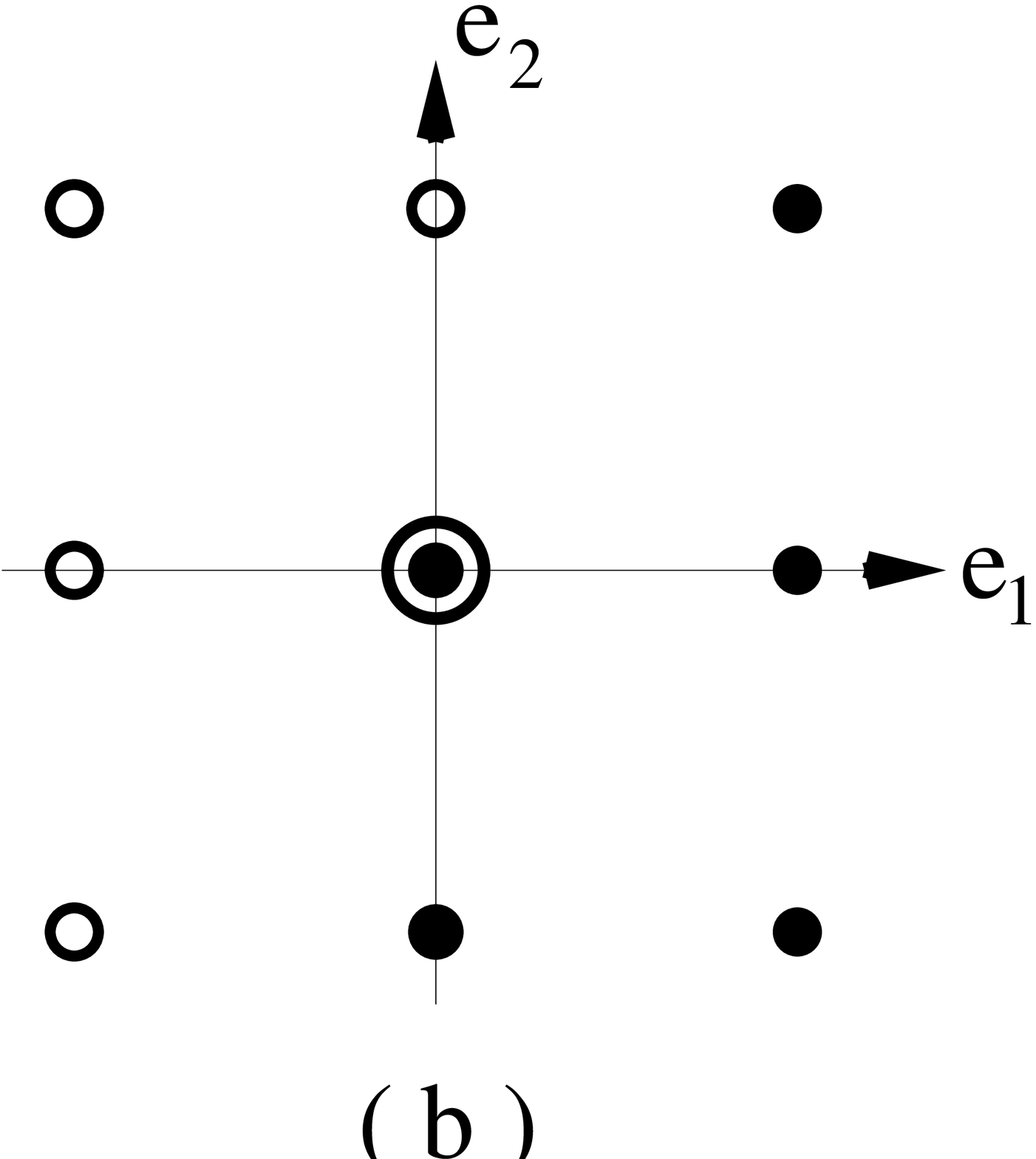} ~
\epsfxsize=1.5in\epsfbox{
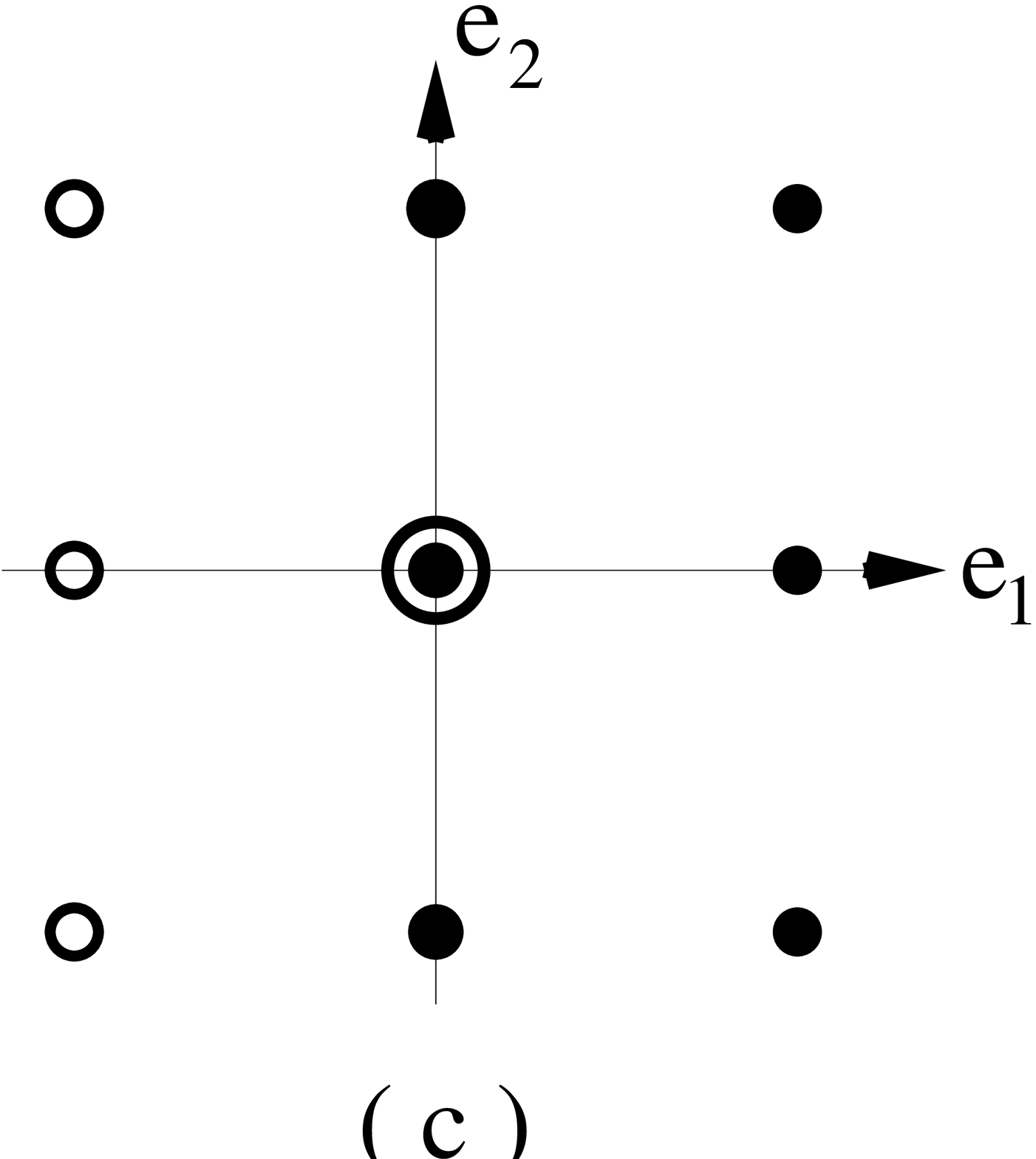}
}
\caption[Root space]{Root space of the complexified conformal Lie algebra
$\mathfrak{conf}_3$, indicated by the full and the open points. The double
circle in the center denotes the Cartan subalgebra. The generators belonging 
to the three non-isomorphic parabolic subalgebras \protect{\cite{Henk03}} 
are indicated by the full points, namely
(a) $\wit{\mathfrak{sch}}_1$, (b) $\wit{\mathfrak{age}}_1$ and
(c) $\wit{\mathfrak{alt}}_1$. 
\label{Bild6}}
\end{figure}

In order to derive (\ref{7:gl:RR}), consider the diffusion equation
\BEQ \label{7:gl:Sch-Gl}
\left(2{\cal M}\partial_t - 
\vec{\partial}_{\vec{r}}\cdot\vec{\partial}_{\vec{r}}\right)
\phi(t,\vec{r}) =0
\EEQ
For fixed $\cal M$, the Schr\"odinger group is the maximal invariance group
on the space of solutions of eq.~(\ref{7:gl:Sch-Gl}). It is defined by the
space-time transformations ($\cal R$ is a rotation matrix)
\BEQ \label{7:gl:SCH}
t \longmapsto t' = \frac{\alpha t + \beta}{\gamma t + \delta} \;\; , \;\;
\vec{r} \longmapsto \vec{r}' = \frac{{\cal R} \vec{r} + \vec{v} t + \vec{a}}
{\gamma t + \delta} \;\; ; \;\; \alpha\delta - \beta\gamma =1
\EEQ
and acts projectively on the solutions $\phi(t,\vec{r})$ \cite{Nied72}. 
Let $\mathfrak{sch}_d$ be the Lie algebra of (\ref{7:gl:SCH}). 
Time-translations occur in $\mathfrak{sch}_d$ and are parametrized by 
$\beta$. If we treat the `mass' $\cal M$ not as a constant but as another 
variable, the embedding $\mathfrak{sch}_d\subset\mathfrak{conf}_{d+2}$ 
for the complexified Lie algebras follows \cite{Henk03}, 
where $\mathfrak{conf}_{d+2}$ is the Lie 
algebra of the conformal group in $d+2$ dimensions. From the classification of 
the parabolic subalgebras of $\mathfrak{conf}_{d+2}$ we obtain several
new subalgebras, called $\wit{\mathfrak{age}}$ or 
$\wit{\mathfrak{alt}}$ \cite{Henk03}. For the $1D$ case, 
we illustrate in figure~\ref{Bild6} 
their definition through the root space of $\mathfrak{conf}_3\cong B_2$. 
These subalgebras still contain the generator
for the dilatations $t\to b^2 t, \vec{r}\to b \vec{r}$ (which is in the
Cartan subalgebra of $\mathfrak{conf}_3$) but do not contain 
time-translations anymore (which is in the lower left corner of 
figure~\ref{Bild6}abc). They are candidates for a dynamic symmetry algebra of
ageing systems. If we assume that the two-time response function transforms
covariantly under the action of either $\wit{\mathfrak{age}}$ or 
$\wit{\mathfrak{alt}}$, a set of linear differential equations for 
$R(t,s;\vec{r})$ is obtained. Matching their solution with the 
expected \cite{Godr02} scaling behaviour of $R$, we recover eq.~(\ref{7:gl:RR})
in the special case $z=2$.

%%----------------------------------------------------------------------------%%
%%----------------------------------------------------------------------------%%

The functional form of $R$ depends on the fact that the Galilei transformation
of (\ref{7:gl:SCH}) is identical to the well-known one of a free particle.
It is not trivial at all that the response function of an interacting
field theory such as the Glauber-Ising model in $d>1$ dimensions should be
recovered from a dynamical symmetry of the equation of motion of a free-field
theory. 

There exist infinite-dimensional Lie algebras which contain $\mathfrak{sch}_d$
as subalgebras. For example, the Schr\"odinger group (\ref{7:gl:SCH}) is a
subgroup of the group defined by the transformations $t\to t'$ and 
$\vec{r}\to\vec{r}'$ where 
\BEQ
t' = \beta(t) \;\; , \;\; \vec{r}' = \vec{r} \sqrt{\dot{\beta}(t)\:}
\;\; \mbox{\rm ~~ or else ~~}
t' = t \;\; , \;\; \vec{r}' = \vec{r} - \vec{\alpha}(t) 
\EEQ
and $\beta$ and $\vec{\alpha}$ are arbitrary functions. 
Whether this has a bearing on the ageing behaviour of non-equilibrium 
spin-systems is still open. Local scale transformations generalizing the
Schr\"odinger group (\ref{7:gl:SCH}) to general values of the dynamical 
exponent $z\ne 2$ exist \cite{Henk02}.
It can be shown that $R(t,s;\vec{r})=R(t,s;\vec{0})\Phi(r (t-s)^{-1/z})$,
such that eq.~(\ref{7:gl:RR}) holds for the autoresponse $R(t,s;\vec{0})$ if
$\lambda_R/2$ is replaced by $\lambda_R/z$ and $\Phi(v)$ is given as the
solution of a linear differential equation of fractional order \cite{Henk02}.

\zeile{1}
\noindent 
I thank M. Pleimling and J. Unterberger for their pleasant collaboration on 
local scale invariance. 
This work was supported by CINES Montpellier (projet pmn2095) and the 
Bayerisch-Franz\"osisches Hochschulzentrum (BFHZ).

\end{document}